\documentclass[a4paper,11pt]{article}
\pdfoutput=1 

\usepackage{jinstpub} 
\usepackage{siunitx}
\usepackage{float}
\usepackage{subfig}
\usepackage{booktabs}
\usepackage{multirow}
\usepackage{pgffor}
\usepackage{amsmath}

\usepackage{lineno}

\newcommand\SNPMT{BA0794}

\newcommand\pmttype{R15458-02}

\title{Homogeneity of the Photocathode in the Hamamatsu \pmttype\ Photomultiplier Tube}%
\author[1]{M.~A.~Unland~Elorrieta,%
\note{Corresponding author.}}
\author{R.~S.~Busse,}
\author{L.~Classen}
\author{and \mbox{A. Kappes}}
\affiliation{Institut f\"ur Kernphysik, Westf\"alische Wilhelms-Universit\"at M\"unster \\ Wilhelm-Klemm-Str. 9, M\"unster, Germany}

\emailAdd{m.unland@wwu.de}

\abstract{
It is common practice to test the optical properties of photomultiplier tubes (PMTs) by illuminating the entire photocathode region from the front at once and measuring the average performance. However, for optimal utilisation of the PMT performance in experiments, especially in the single-photon region, it is essential to also know the systematic variations across the photocathode, which requires measurements with focused light sources that illuminate only small regions of the PMT. We present a detailed uniformity characterisation of the gain, transit time, transit time spread, and pulse shape of the 80$\,$mm Hamamatsu R15458-02 PMT. We find that the parameters exhibit asymmetry along one axis, likely caused by the position and geometry of the dynode system. For all parameters except the transit time, the observed variations are small given the intrinsic variation of the parameters. For positions with shifted transit time we observe on average underamplified pulses which can potentially be exploited to improve the pulse reconstruction.
}

\keywords{Photon detectors for UV, visible and IR photons (vacuum); Cherenkov detectors; Large detector systems for particle and astroparticle physics}

\begin{document}
\flushbottom
\maketitle
\section{\label{sec:introduction}Introduction}
With KM3NeT \cite{Adri_n_Mart_nez_2016} (under construction in the Mediterranean Sea), IceCube-Upgrade \cite{ishihara2019icecube} (to be constructed at the South Pole) and Hyper-Kamiokande \cite{Abe:2018uyc} (to be constructed in Japan),
so-called multi-PMT optical modules \cite{nim:a718:513, Classen:2019/w, DeRosa:2020wbm} have been designed for three next-generation neutrino detectors. At the heart of this innovation in optical module technology is the use of several relatively small PMTs inside a single transparent pressure vessel instead of one large PMT. Among other advantages, the design provides a large sensitive area, relatively homogeneous sensitivity over the entire solid angle, and intrinsic directional information about the recorded photons.
\begin{figure}[b]
\centering
 \includegraphics[width=0.5\textwidth]{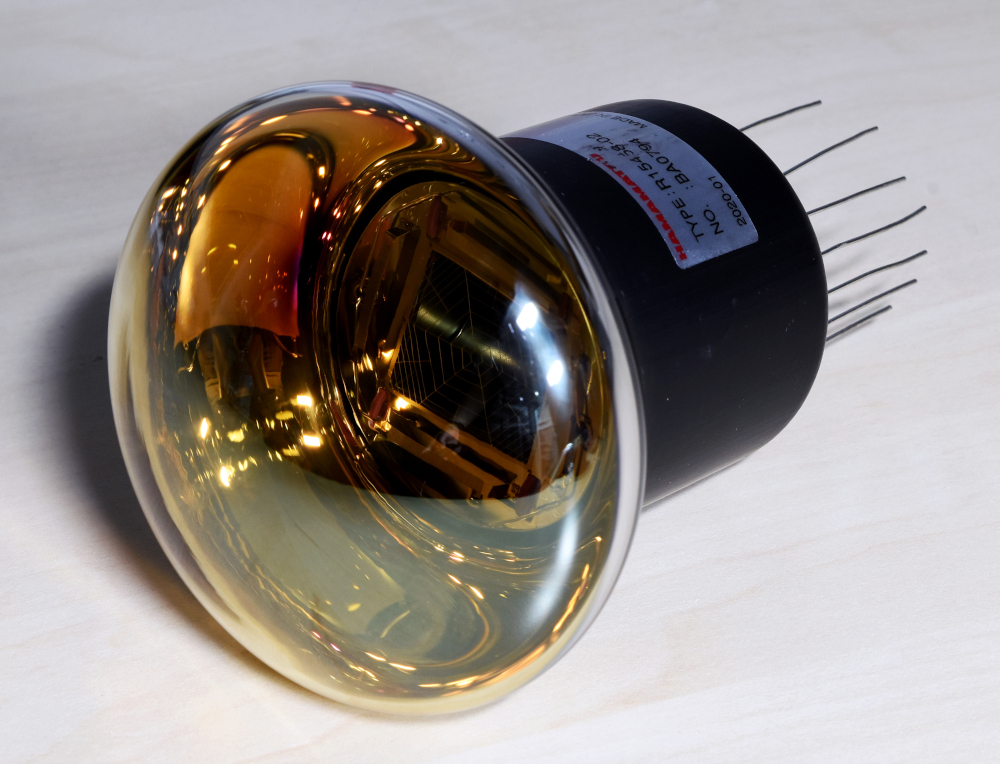}
 \caption{Picture of the $\SI{80}{mm}$ \pmttype\ PMT.}
\label{fig:PMTpic}
\end{figure}

 The Hamamatsu PMT type \pmttype\ (figure~\ref{fig:PMTpic}) was developed for the special conditions found in IceCube and is the successor to the PMT type R12199-01 HA MOD. The latter was based on the PMT type R12199-02, developed by Hamamatsu for the KM3NeT experiment from the commercial flat window PMT type R6233. The size of the IceCube module is limited by the diameter of the ice boreholes which requires short PMTs. Hence, the tube length was reduced from $\SI{98}{mm}$ to $\SI{93}{mm}$ resulting in R12199-01 HA MOD. 

 In order to reduce and stabilise the dark noise rate of the PMT operating at negative high voltage, a conductive layer at photocathode potential covering the outer surface of the tube (``HA coating'') was added to the design. The main characteristics of the R12199-01 HA MOD were found to be similar to those of its predecessor and suitable for low-temperature applications \cite{UnlandElorrieta:2019yhd}.
 
For the Hamamatsu \pmttype\ PMT, the tube length was further reduced to \SI{91}{mm}. Its performance is expected to be similar to that of R12199-01 HA MOD. Therefore, the results shown in this work should be applicable to both types of \SI{80}{mm} PMTs used in KM3NeT and IceCube-Upgrade.

It is common practice to use frontal light illumination (a plane wave perpendicular to the PMT axis) for acceptance and characterisation tests of PMTs, thus obtaining average performance parameters. This procedure assumes that the photocathode is relatively homogeneous for these parameters over its entire surface. If this is not the case, the distributions of the parameter values depend on the angle of incidence of the plane wave. In this work we investigate the uniformity of these parameters over the whole sensitive area by means of an elaborate measurement technique. Parameters investigated are the gain, the transit time and the transit time spread (TTS) as well as the pulse shape parameters pulse width, rise time and fall time.

\section{\label{sec:methods}Experimental setup and measurement method}
\begin{figure}[tb]
\centering
 \includegraphics[trim=0 8cm 7cm 0, clip, width=0.7\textwidth]{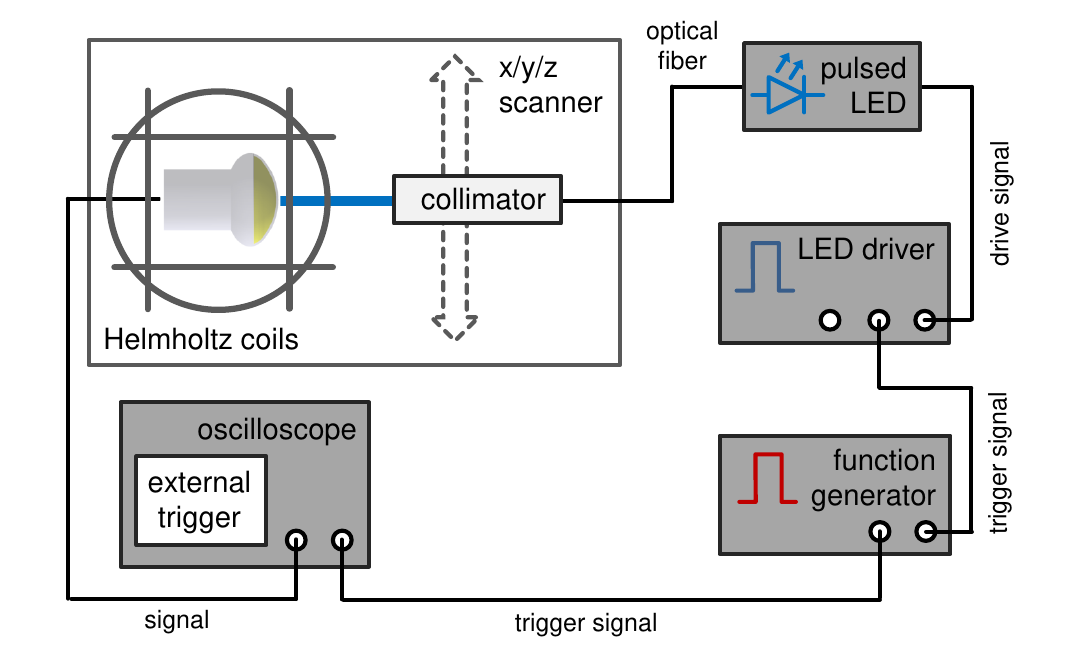}
 \caption{Schematic layout of the parameter scan setup.}
\label{fig:scan_setup}
\end{figure}

The uniformity of the response across the photocathode was studied using the setup depicted in figure~\ref{fig:scan_setup}. Small areas of the photocathode were illuminated by pulses of light from an LED\footnote{PLS-8-2-719 controlled by PicoQuant PDL 800-B, wavelength $\SI{459}{nm}$ (FWHM $\SI{24}{nm}$).} connected to a $\SI{105}{\micro \meter}$ multi-mode optical fibre. The LED was triggered by a function generator\footnote{RIGOL DG1032Z} with a repetition rate of $\SI{30}{kHz}$. Upon exiting the fibre, the light was collimated\footnote{60FC-SMA-0-A7.5-01 lens collimator by Sch\"after \& Kirchhoff, collimated at $\SI{532}{nm}$}, resulting in a light beam with a FWHM of $\sim \SI{0.9}{mm}$ at a distance of $\SI{60}{mm}$ from the collimator. 
The fibre was attached to a 3D scanner with which it was moved in x/y/z direction. The precision on the position of the fibre is better than $\SI{50}{\micro m}$.
The frontal curvature of the photocathode was compensated by adjusting the distance between the illuminated surface and the fibre for each measurement, so that the photocathode was always illuminated at the focal length with a constant beam diameter.
 
The measurement was performed inside three Helmholtz coils compensating the for the Earth's magnetic field. The custom-built coils produce a field with relative  deviations of less than $1\%$ over the entire volume of the PMT. This ensures that magnetic field inhomogeneities do not affect the measurements and renders the results independent of the position and orientation of the PMT.
\begin{figure}[t]
\centering
\includegraphics[scale=1]{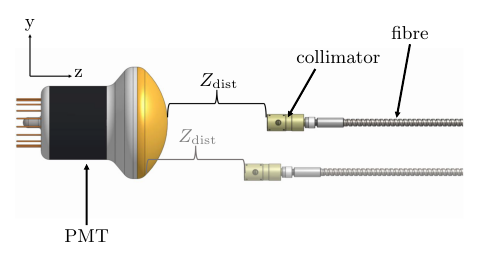}
\hfill
\includegraphics[scale=1]{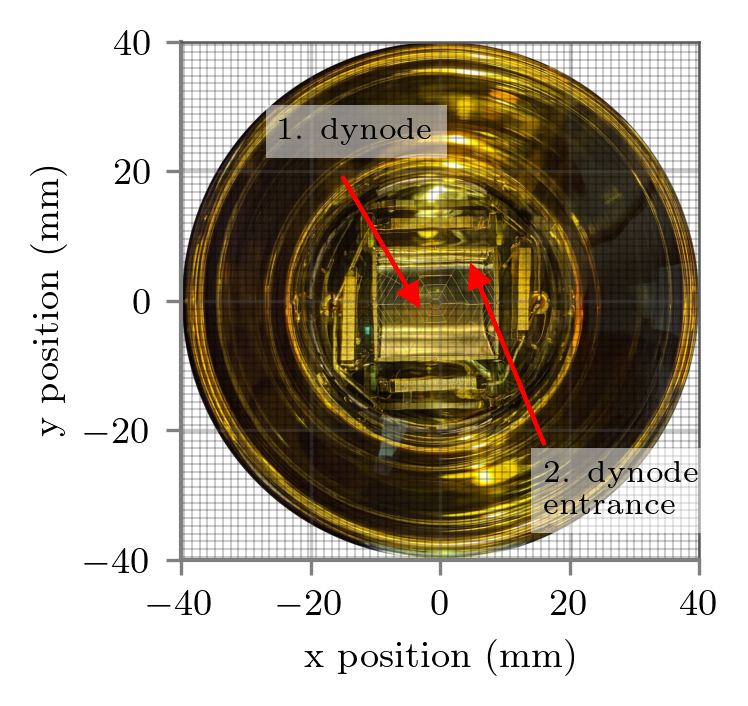}
\caption{\emph{Left}: Schematic of the positioning of the PMT and optical fibre. The photocathode was always illuminated perpendicular to the tube axis (z), and the distance $Z_{\rm{dist}}$ between its surface and the collimator is constant. \emph{Right}: Frontal picture of the PMT. The centre of each grid box represents the location of the measured points on the photocathode. The position and scale of the PMT picture is approximated and may not correspond to reality.}
\label{fig:expsetup:grid}
\end{figure}

The photocathode was scanned with the light always parallel to the axis of the tube (figure~\ref{fig:expsetup:grid} left) on a grid of $\SI{1.2}{mm} \times \SI{1.2}{mm}$. The coordinate system used in this work is shown on the right side of figure~\ref{fig:expsetup:grid}. After the trigger of the function generator, the anode signal of the PMT (waveform) was digitised with an oscilloscope\footnote{PicoScope 6404C} and then directly analysed with a computer. Here, the voltage and arrival time of the minimum of the waveform were stored and the portion within $\pm \SI{15}{ns}$ around the minimum was integrated. This results in a pulse charge in Weber, which is converted to Coulomb by dividing it by the oscilloscope's input resistance of $\SI{50}{\ohm}$. At each grid position, \SI{45000} waveforms were measured.

\begin{figure}[t]
\centering
\includegraphics[scale=1]{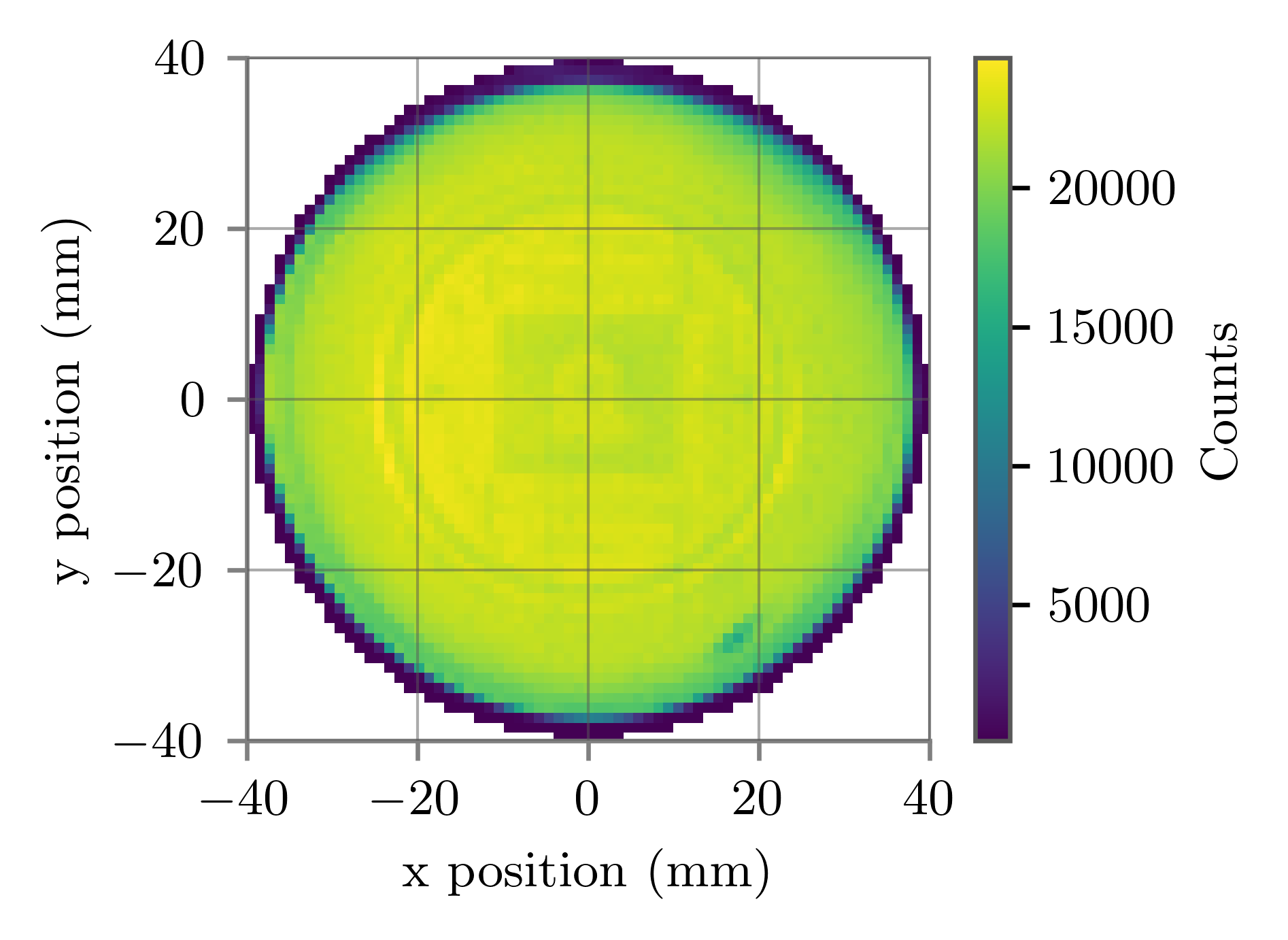}
\hfill
\includegraphics[scale=1]{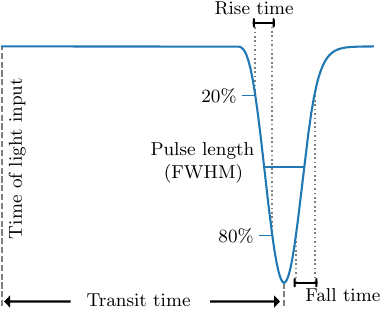}
\caption{\emph{Left}: Number of measured pulses at all photocathode points of PMT \SNPMT. \emph{Right}: Sketch of shape and timing parameters of a PMT signal. 
}
\label{fig:expsetup:counts}
\end{figure}

The shape of a PMT pulse can be characterised by its length (FWHM) and by its rise and fall times, the latter defined by the times required for the signal to cross two successive voltage thresholds. The most common definition is using the $\SI{10}{\percent}$ and $\SI{90}{\percent}$ of the pulse height. However, since the average pulse amplitude is between 7 to $\SI{10}{mV}$, the $\SI{10}{\percent}$ level lies often inside the baseline noise of the oscilloscope, especially in low amplitude pulses. Therefore, the threshold levels chosen in this work were $\SI{20}{\percent}$ and $\SI{80}{\percent}$ of the pulse amplitude as depicted on the right side of figure~\ref{fig:expsetup:counts}. The crossing times were estimated by interpolating between the sample points of the waveform. 

As depicted in the right plot of figure~\ref{fig:expsetup:counts}, the transit time of the PMT is the time between the photon absorption in the photocathode and the arrival of the pulse. Since in this setup the absolute time of the light emission is unknown, in this work the transit time is always presented relative to a reference time. The transit time spread (TTS) is commonly defined either as the standard deviation or the FWHM of the transit time distribution. In the following, the TTS will refer to the former, the standard deviation. Since the arrival time of the pulse is taken at the pulse minimum, variations of the pulse shape (pulse length, rise time and fall time) have only a small impact on the calculated TTS.

The scan was performed in the areas defined by $\sqrt{x^2+y^2}<\SI{41}{mm}$ which is larger than the projected area of the photocathode ($r=\SI{40}{mm}$).  This ensures that the entire front of the PMT is measured. In the analysis, only grid points with a sufficient number of measured pulses were considered. The minimum count chosen was $\SI{10}{\percent}$ of the average number of pulses measured in the central region of the PMT ($r<\SI{30}{mm}$). On the left side of figure~\ref{fig:expsetup:counts}, the number of pulses measured for each grid point from an exemplary measurement is depicted. By comparing it with the frontal image of the PMT (see right side of figure~\ref{fig:expsetup:grid}), many of the internal elements of the PMT can be identified. This is due to reflections of light from these elements, which allows multiple crossing of the beam with the photocathode. 

To compare the result with frontal plane wave measurements, the fibre exit was placed at a distance of $\SI{1.5}{m}$ from the PMT after each scan and the collimator was replaced by a diffuser, which illuminates the entire photocathode. Afterwards, \SI{500000} waveforms were measured and analysed in the same way as the scan data.

Altogether, five PMTs were measured with the described procedure. However, in the following sections, only the results of the PMT with serial number \SNPMT\ are shown in detail. The results of the other PMTs are comparable.

\section{\label{sec:pulse}Uniformity of the photocathode with regard to the pulse characteristics}
In this chapter, the photocathode uniformities for different pulse parameters are shown. In Section \ref{sec:charge} we present the homogeneity for the gain, followed by the homogeneity for transit time and transit time spread in Section~\ref{sec:timing}. Finally, in Section \ref{sec:shape}, the uniformity of the pulse shape parameters ---pulse width, rise time and fall time--- is shown.

\subsection{\label{sec:charge}Gain}

\begin{figure}[tb]
\centering
\includegraphics[scale=0.95]{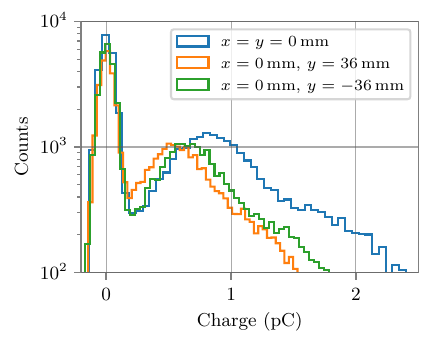}
\hfill
\includegraphics[scale=0.97]{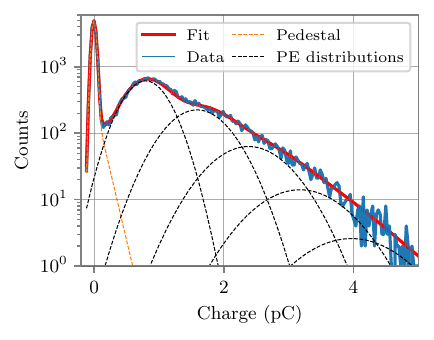}
\caption{\emph{Left}: Charge histogram of three different positions (s. figure~\ref{fig:expsetup:grid}) on the photocathode. \emph{Right}: Charge histogram fitted with the function~\ref{eq:GainFit} in red. The orange dashed line marks the sum
of the pedestal and the tail caused by underamplified pulses. The black dashed lines represent the contribution from single to five PEs.}
\label{fig:gain:singledistributions}
\end{figure}

The charge distribution of the waveforms for a point on the photocathode can be plotted as a histogram as shown in the figure~\ref{fig:gain:singledistributions}. From an LED light pulse, the PMT will record a discrete number of photons following Poisson statistics. The measured charge for the case where no photons from the LED were recorded by the PMT is represented in the histogram as a Gaussian peak and is called pedestal. The shape of the pedestal depends on the amplitude distribution of the baseline noise and the rate of background events. The charge of events in which single photoelectrons (SPE) are emitted forms a broad peak around the mean SPE charge $Q_1$ to the right of the pedestal. The width of this distribution $\sigma_1$ is given by the charge resolution of the PMT and is usually too large to separate SPE events from those with multiple photoelectrons (MPE), but narrow enough to distinguish them from the pedestal. MPE events form even wider distributions next to the SPE peak. The charge distribution of three different points on the photocathode of the PMT with serial number \SNPMT\ is shown on the left side of figure~\ref{fig:gain:singledistributions}. It is noticeable that the position of the SPE peak changes along the photocathode, which means that different points on the photocathode have different \textit{local} gains. The gain in this case corresponds to $G={Q_1}/{e}$, where $e$ is the elementary charge. To determine $Q_1$, it is necessary to fit the charge distributions.

The model describing PMT charge distributions used in this work is based on \cite{Bellamy1994}. Here, the charge output of a PMT is described as a convolution between three functions:
\begin{itemize}
    \item a Poisson distribution modelling the number of detected photons per LED pulse $n$, \\$P(n|\mu)=\mu^n \cdot \textrm{e}^{-\mu}/n!$, where $\mu$ is the mean number of detected photoelectrons;
    \item the ``ideal'' PMT spectrum, which describes the SPE PMT response without background. This is modelled by a Gaussian with mean $Q_1$ and standard deviation $\sigma_1$;
    \item a background function, which is the sum of a Gaussian (with mean $Q_0$ and standard deviation $\sigma_0$), describing the pedestal, and an exponential distribution $\lambda \cdot \textrm{e}^{-\lambda q}$. The latter models the contribution from underamplified pulses (such as prepulses released at the first dynode) and other discrete processes, such as thermionic emission or background light detected in (partial) coincidence with the LED light pulse. Nevertheless, the dark rate at room temperature of this PMT type is less than $\SI{1000}{s^{-1}}$ and the time window for the pulse integration is $\SI{30}{ns}$ in length, therefore, a (partial) coincidence with the LED light pulse is extremely improbable.
\end{itemize}
The convolution of the three functions results in the probability density function (pdf) used in this work:
\label{appendix:charge}

\begin{align}
\label{eq:GainFit}
\begin{split}
     f(q) &= \sum^\infty_{n=0}\frac{\mu^n \cdot \textrm{e}^{-\mu}}{n!}\Big[(1-P_u)\cdot G(q,Q_n,\sigma_n)+ \\ &+P_u \cdot \frac{\lambda}{2} \cdot\textrm{e}^{-\lambda \cdot (q-Q_n-\frac{1}{2}\lambda n\sigma_1^2)}\cdot\left[ 1- \textrm{erf}\left(\frac{Q_n-q+\lambda\cdot n\cdot\sigma^2_1}{\sqrt{2n}\sigma_1}\right)\right]\Big],
     \end{split}
     \end{align}
where $P_u$ the probability of measuring a non-Gaussian charge contribution from background signals and
\begin{equation} 
\begin{split}
Q_n & = Q_0+n\cdot Q_1, \\
\sigma_n & = \sqrt{\sigma_0^2+n\cdot \sigma_1^2}, \\
G(x,\mu,\sigma) & = 1/(\sqrt{2\pi}\sigma)\cdot\exp{(-(x-\mu)^2/2\sigma^2)}.
\end{split}
\end{equation}
The infinite series of equation~\ref{eq:GainFit} was truncated when $(1-\sum^N_{n=0}\frac{\mu^n \cdot \textrm{e}^{-\mu}}{n!})<10^{-4}$. The data were fitted using the binned maximum likelihood method, assuming a Poisson distribution for the bin counts. An example of a charge distribution fit is shown on the right side of figure~\ref{fig:gain:singledistributions}.
\begin{figure}[tb]
\centering
\includegraphics[scale=1]{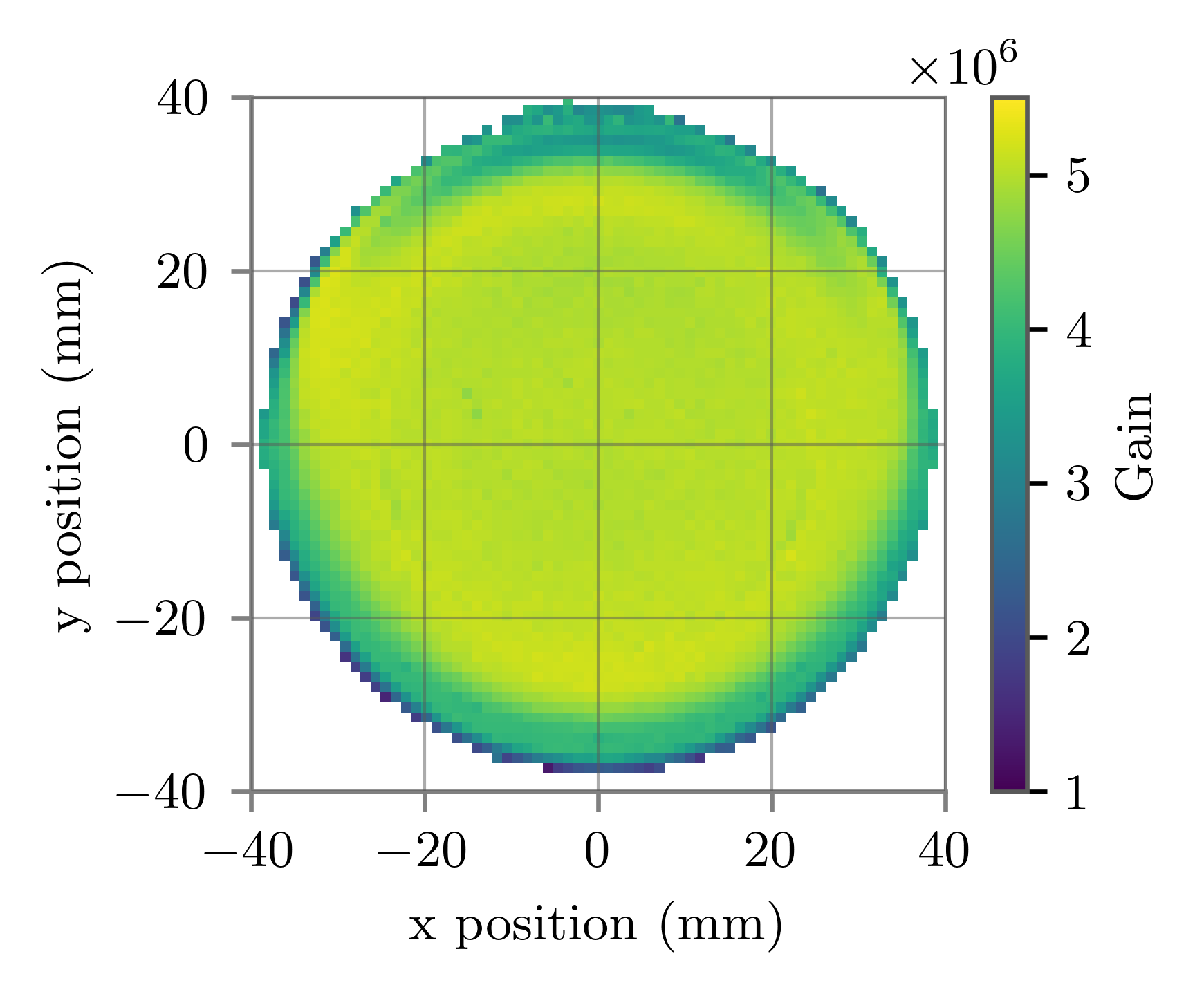}
\hfill
\includegraphics[scale=1]{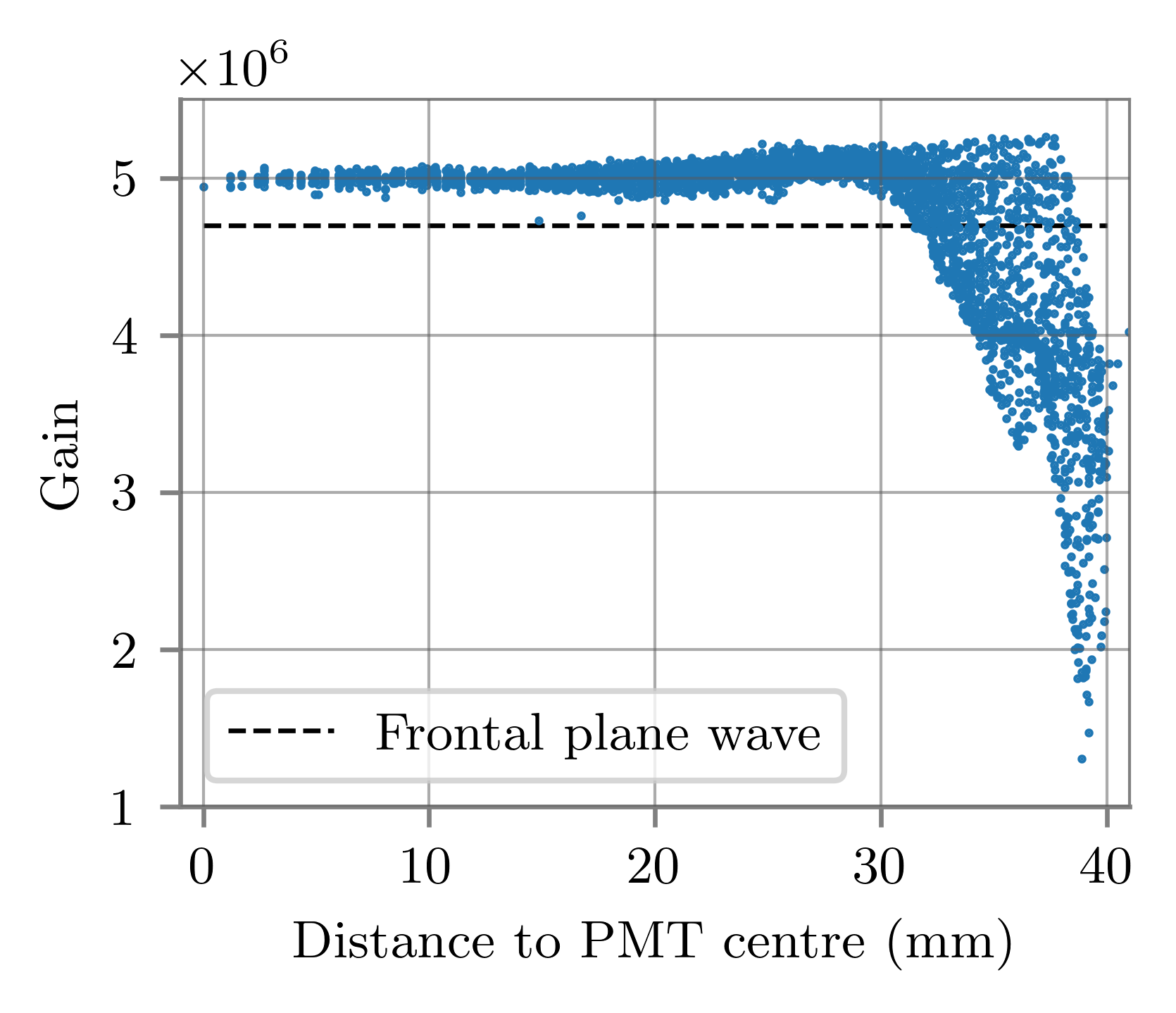}
\caption{\emph{Left}: Gain scan of PMT \SNPMT. \emph{Right}: Local gain values as a function of the distance to the PMT centre.}
\label{fig:gain:map}
\end{figure}

Fitting the charge distribution for all points on the photocathode results in the gain map shown on the left side of figure~\ref{fig:gain:map}. For points less than $\SI{30}{mm}$ from the centre of the PMT, the gain remains fairly constant at around $5\cdot10^6$ (nominal gain specified by manufacturer), as depicted on the right side of figure~\ref{fig:gain:map}. Towards the edges, the deviations increase to larger and smaller values. The largest asymmetry appears with respect to the $x$-axis which probably arises from the geometrical asymmetry of the PMT with respect to the orientation of the first and second dynodes (see figure~\ref{fig:expsetup:grid}). A measurement with frontal light yields a lower gain (dashed line in figure~\ref{fig:gain:map} right) than the local gain for most of the photocathode due to the gain reduction at the edges.

\begin{figure}[tb]
\centering
\includegraphics[scale=0.95]{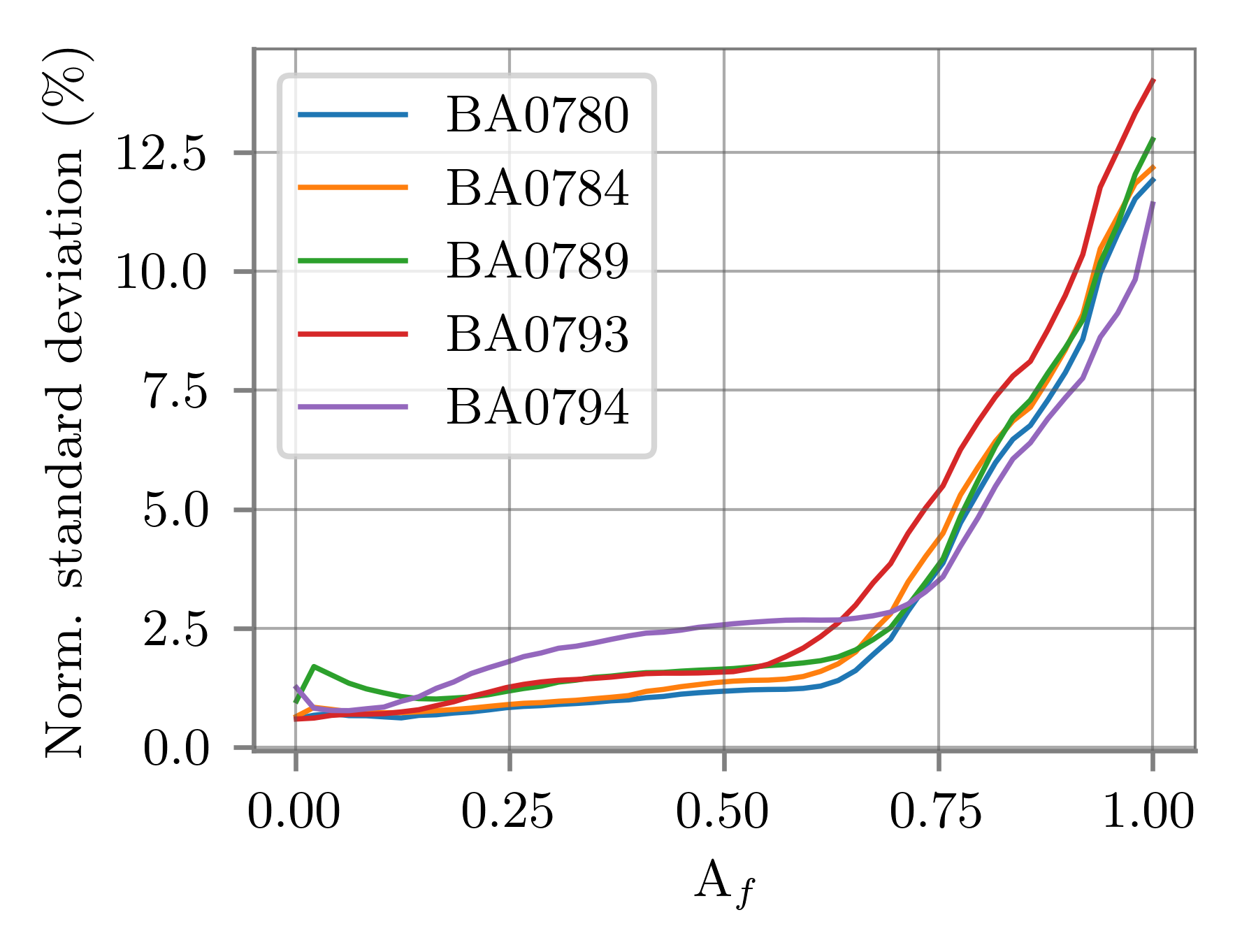}
\caption{Normalised standard deviation of the local gains in $r<\sqrt{A_f}\cdot40\,$mm in dependence of the fraction of area $A_f$ to which the points used correspond. The standard deviation is normalised to the mean gain of the central region $r<30\,$mm. The legend corresponds to the serial number of the measured PMT.}
\label{fig:timing:GainSD}
\end{figure}

As a measure of the homogeneity of the photocathode, the standard deviation of the local gains normalised to the average central gain ($r<30\,$mm) is plotted in figure~\ref{fig:timing:GainSD}. Here, the standard deviation was calculated for all points with $r<\sqrt{A_f}\cdot40\,$mm, where $A_f$ is the fraction of the photocathode area comprising these points. The homogeneity is similar for all PMTs, and the width of the variation remains relatively constant below 2.5\% up to $A_f\sim0.6$. At the edges the standard deviation increases to $\sim\SI{12}{\percent}$. The charge resolution (the width of the SPE charge peak) for this PMT type is on average \SI{46}{\percent} \cite{UnlandElorrieta:2019yhd}. Therefore, at SPE level the effects of gain inhomogeneities would be difficult to distinguish from the intrinsic charge variation.

\subsection{\label{sec:timing}Transit time and transit time spread}
The TTS defines the intrinsic uncertainty of the photon detection time of a PMT. The main contribution to the TTS results from the different paths taken by the photoelectron from the photocathode to the first dynode and by the secondary electrons between the subsequent dynodes \cite{Wright2017}. The PMT photocathode under study is hemispherical to minimise these variations.
\begin{figure}[tb]
\centering
\includegraphics[scale=1]{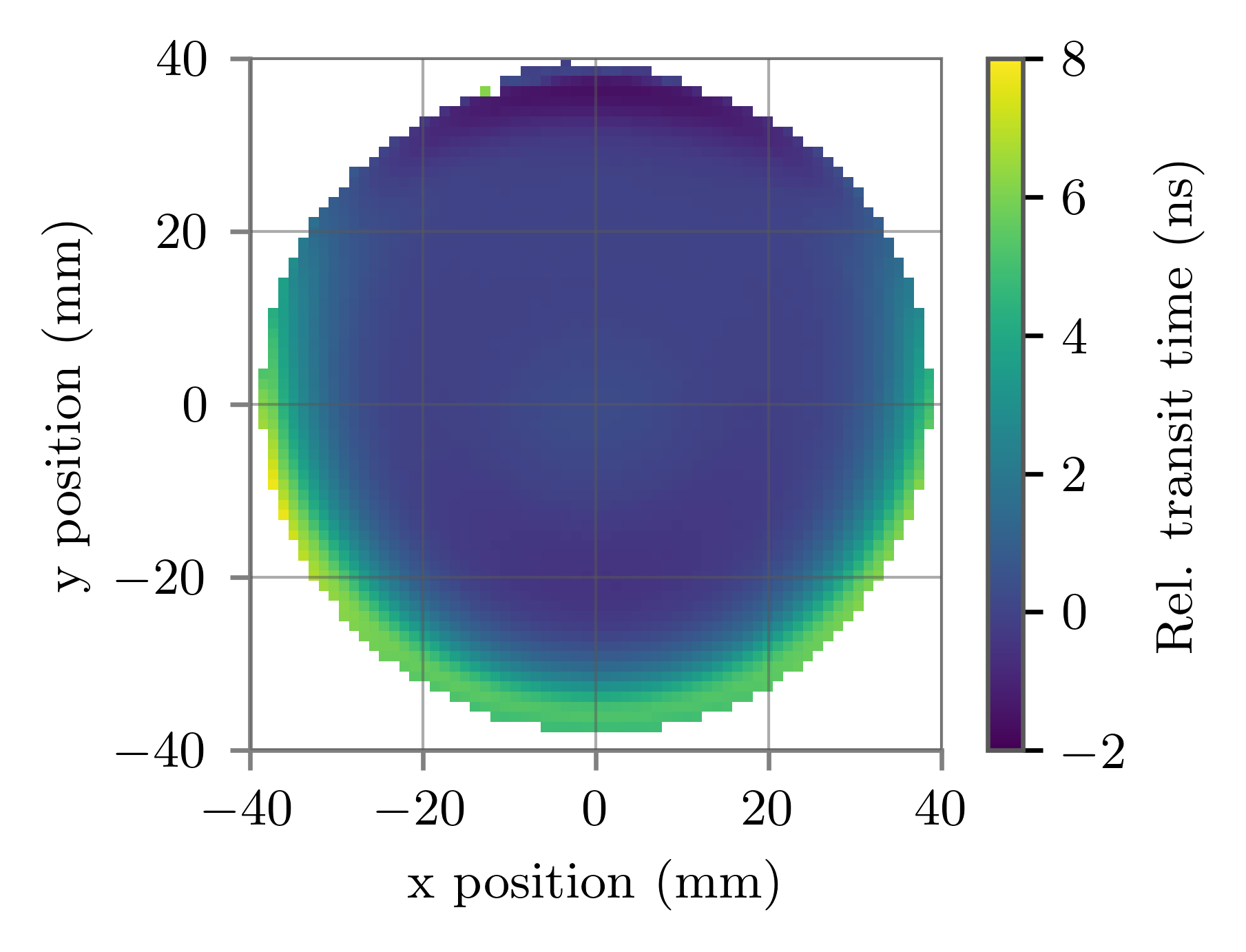}
\hfill
\includegraphics[scale=1]{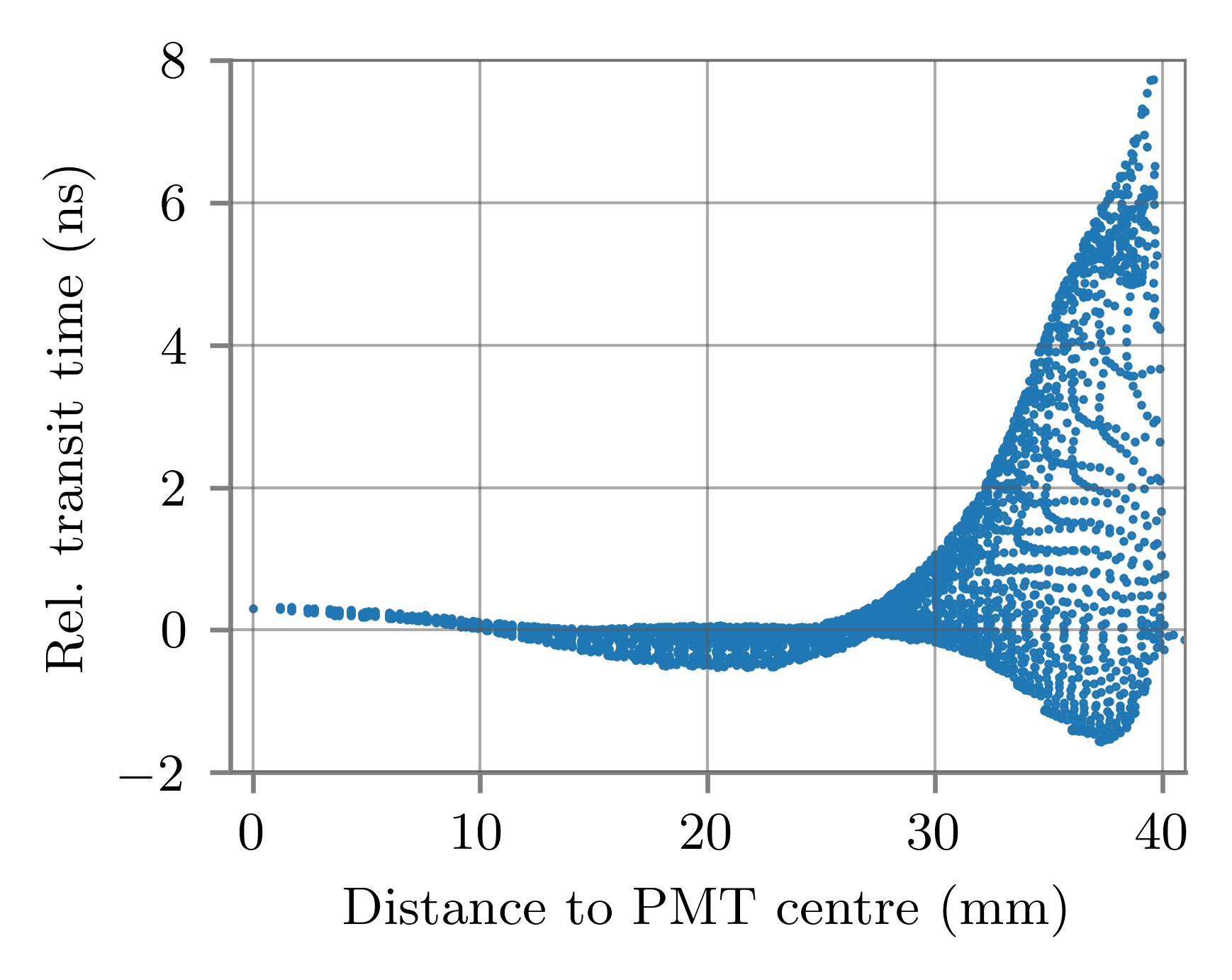}
\caption{\emph{Left}: Photocathode scan of the transit time, relative to the average transit time of the PMT centre ($r<\SI{30}{mm}$). \emph{Right}: Same relative transit time data, but plotted against distance to the PMT centre. Data of PMT \SNPMT.}
\label{fig:timing:TT}
\end{figure}

 To calculate the relative transit time and TTS, the measured arrival times were grouped in a histogram and fitted with a Gaussian. The mean of the Gaussian is the average arrival time  of the signal relative to the trigger and the standard deviation is used for the calculation of the TTS. The mean arrival time for the centre of the PMT ($r<\SI{30}{mm}$) is calculated and subtracted from all fitted averages. The results for the relative transit time are shown in figure~\ref{fig:timing:TT}. There is a noticeable asymmetry along the $y$-axis, going from $\sim\SI{-2}{ns}$ to $\sim\SI{8}{ns}$ edge to edge. Along the $x$-axis the transit time tends to be more symmetric. 

\begin{figure}[tb]
\centering
\includegraphics[scale=0.95]{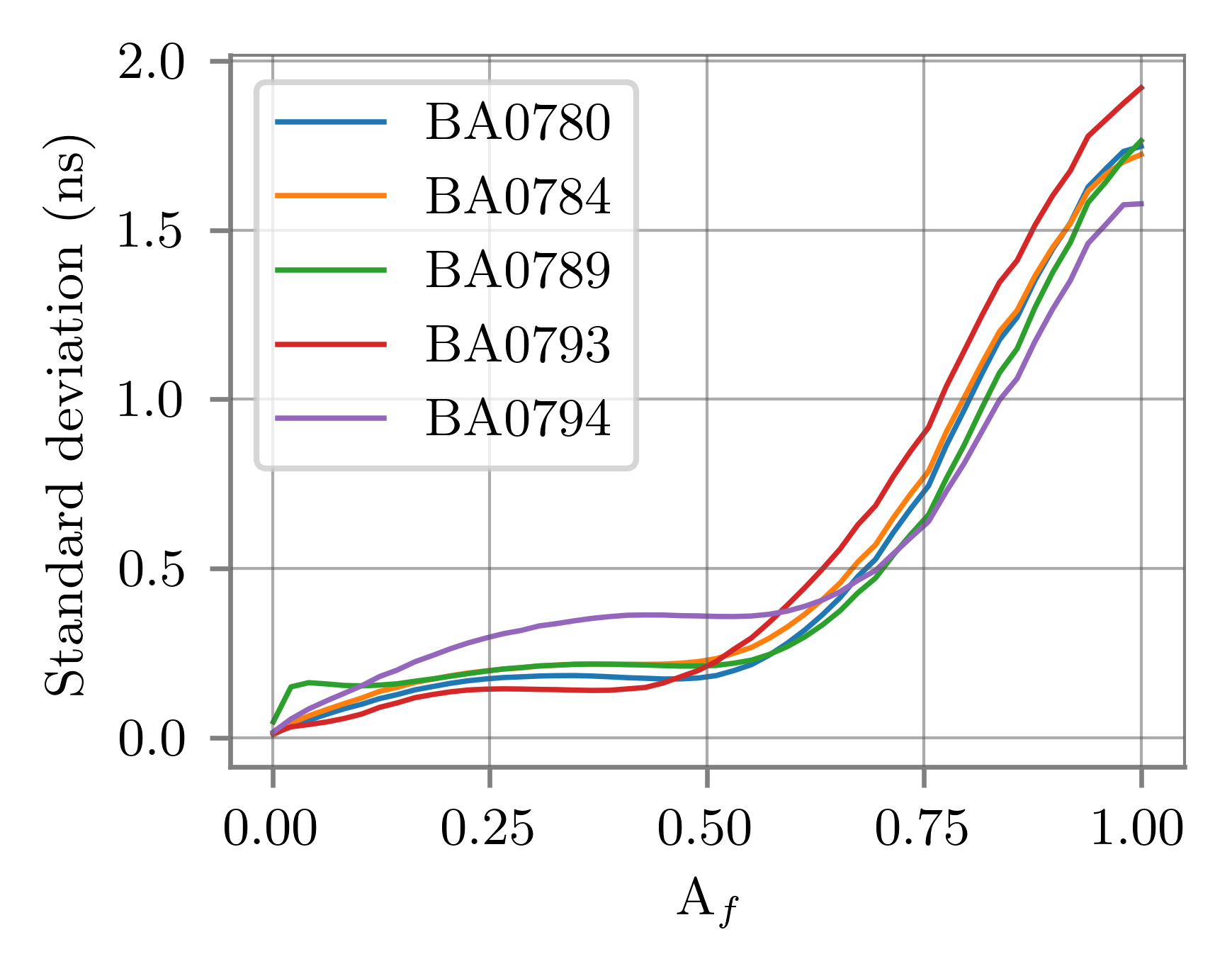}
\hfill
\includegraphics[scale=0.95]{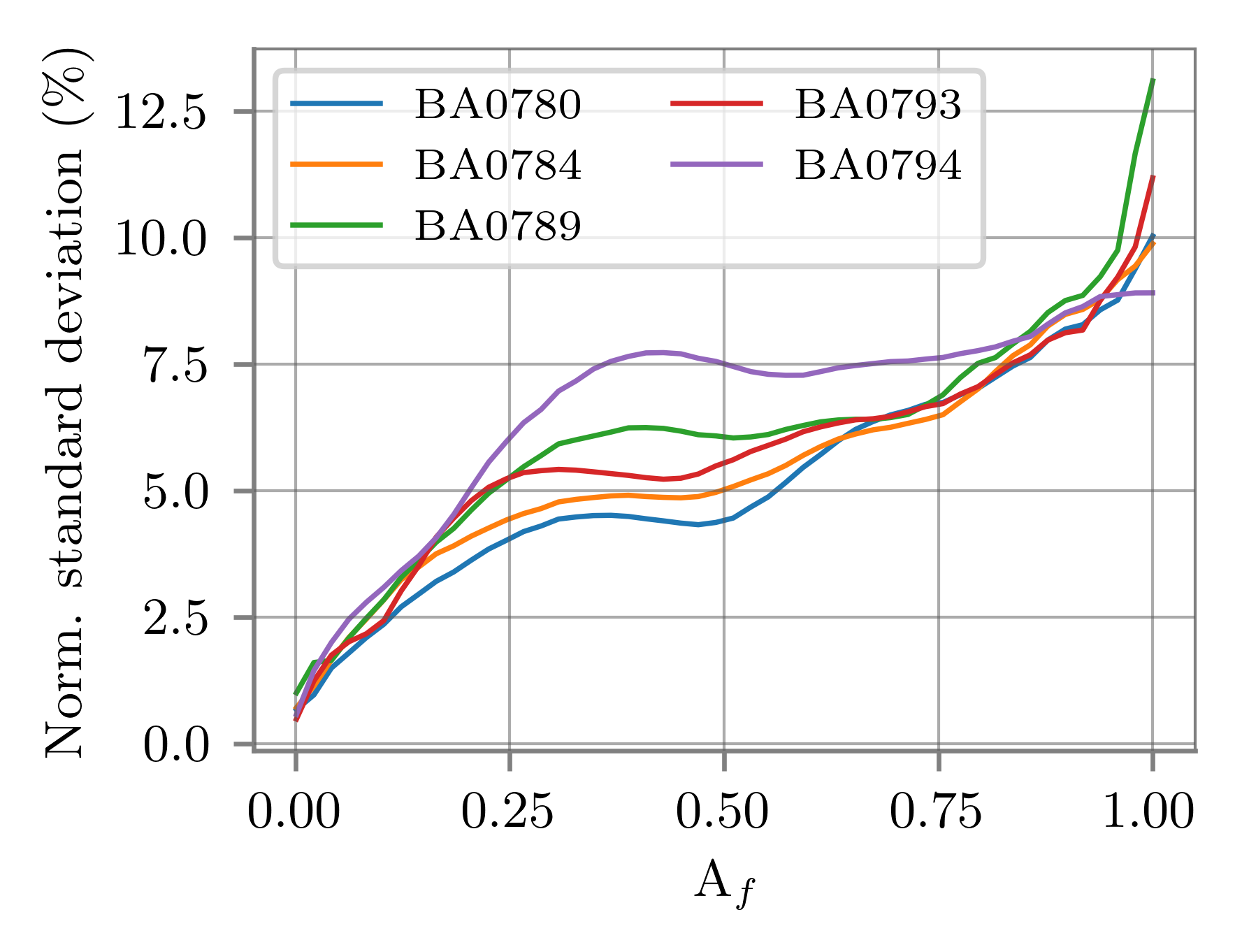}
\caption{\emph{Left}: Standard deviation of the local transit time in $r<\sqrt{A_f}\cdot40\,$mm in dependence of the area fraction $A_f$. \emph{Right}: Standard deviation of the local TTS in $r<\sqrt{A_f}\cdot40\,$mm against the area fraction $A_f$, normalised to the average TTS of the central region $r<30\,$mm.}
\label{fig:timing:TTSD}
\end{figure}

The standard deviation of the local transit time as a function of the area fraction is depicted on the left side of figure~\ref{fig:timing:TTSD}. As with the gain, the standard deviation is relatively flat until $\sim\SI{60}{\percent}$ and increases towards the edges. The standard deviations for the total photocathode area range from $\sim\SI{1.6}{ns}$ to $\sim\SI{1.9}{ns}$, depending on the PMT, putting their variation in the same order of magnitude as the TTS.

\begin{figure}[tb]
\centering
\includegraphics[scale=0.95]{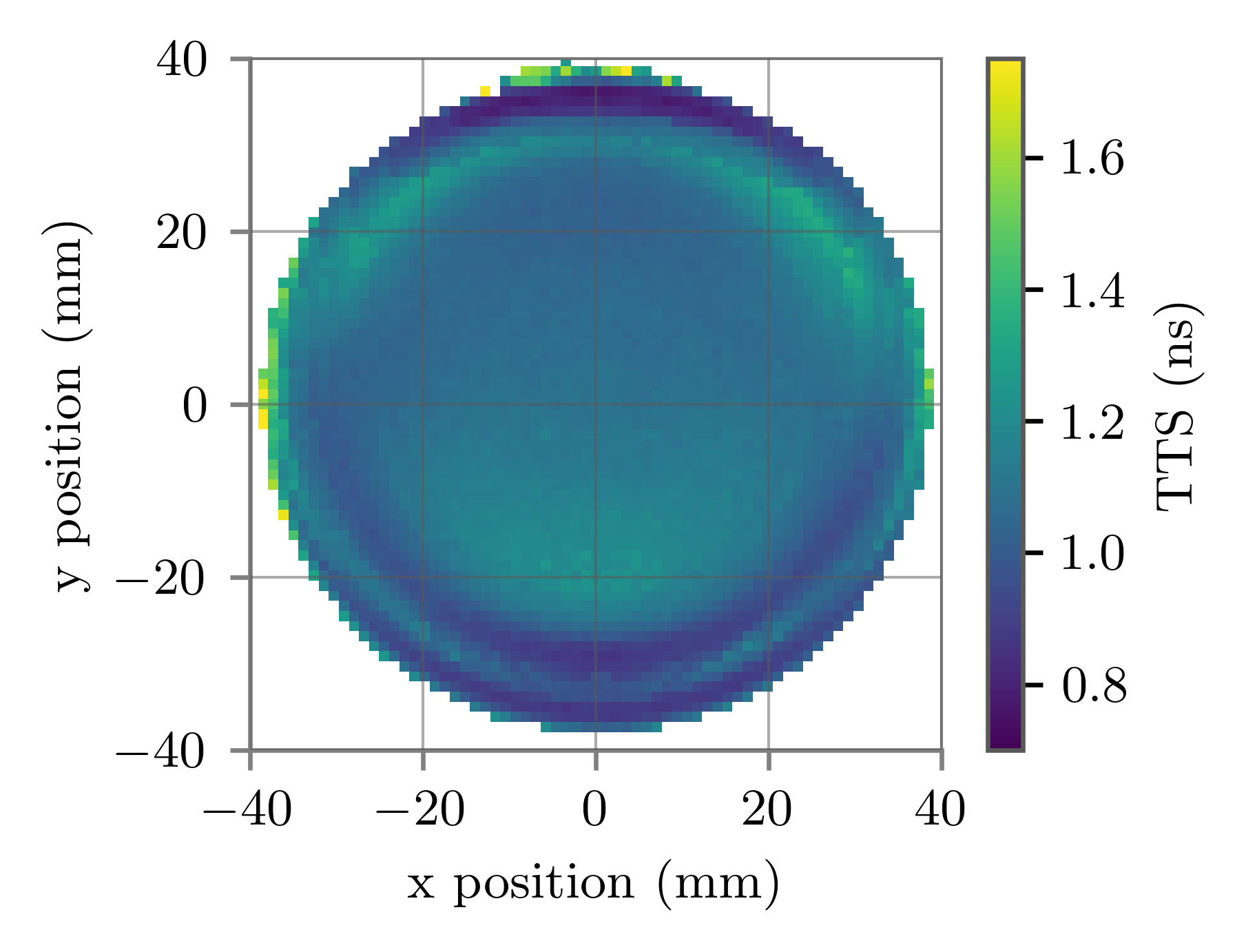}
\hfill
\includegraphics[scale=0.95]{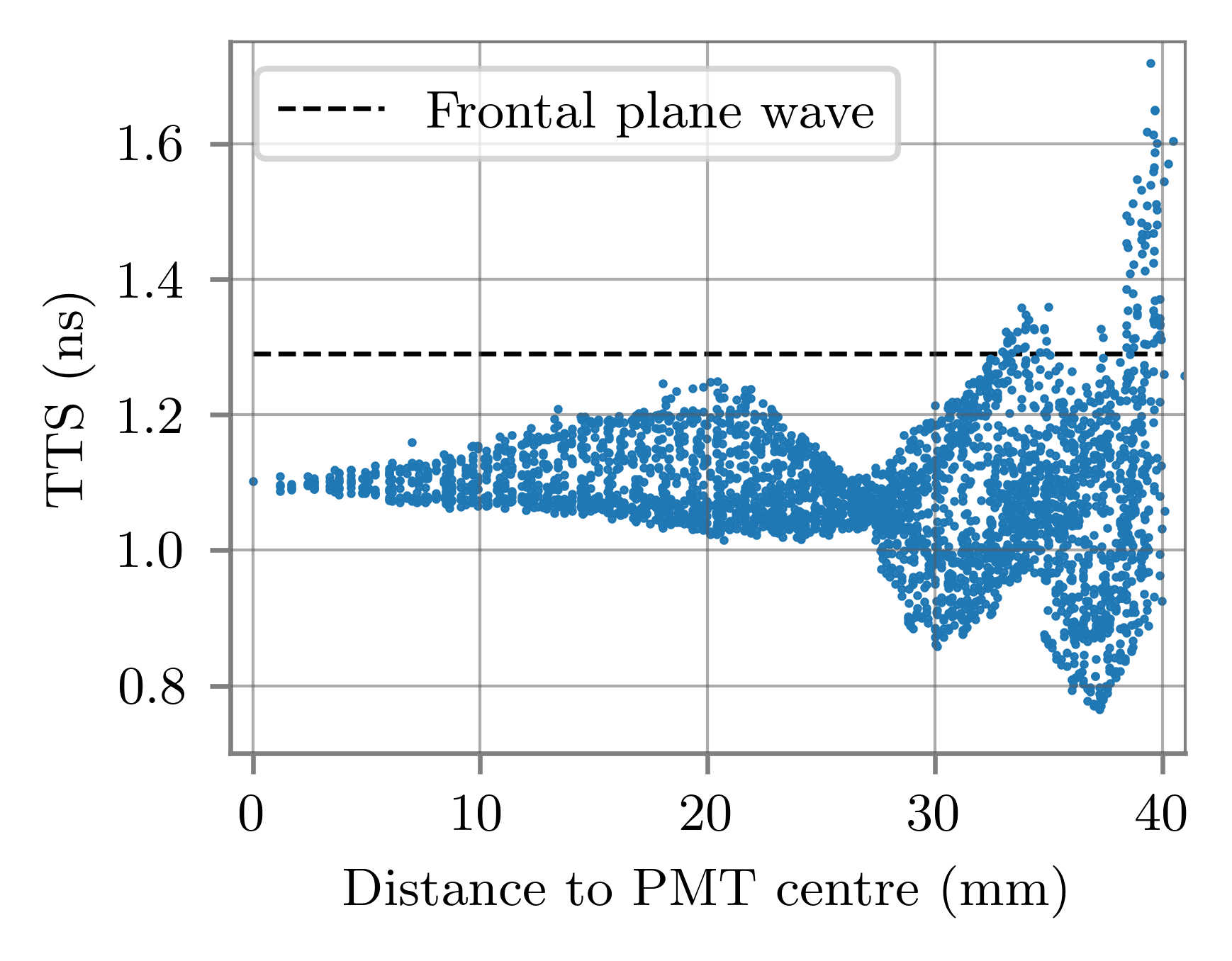}
\caption{\emph{Left}: Transit time spread scan. \emph{Right}: Transit time spread plotted against the distance to the PMT centre. Data of PMT \SNPMT.}
\label{fig:timing:TTS}
\end{figure}

To calculate the TTS, the jitter of the LED and the trigger must first be subtracted from the measured standard deviation. The result has to be further corrected for the fact that some pulses are MPE. More details on these corrections are presented in Appendix \ref{appendix:TTscorrection}. The TTS map of the PMT \SNPMT\ is shown on the left side of figure~\ref{fig:timing:TTS}. On the right side the TTS is plotted against the distance to the PMT centre. The TTS map shows a similar asymmetry along the $y$-axis as the gain and relative transit time scan, but changes more abruptly with $r$, as can be seen on the right of figure~\ref{fig:timing:TTS}. Furthermore, the standard deviation of the TTS points on the right side of figure~\ref{fig:timing:TTSD} shows a different behaviour than that of transit time and gain, with a fairly steady increase of the standard deviation towards the edges of the PMT. When the light beam reflects off the internal structures of the PMT, it does not have to return to the same point of the photocathode where it entered. If the reflected beam hits a point on the photocathode that has a very different transit time than the entry position, the time distribution is widened. This means, that the scan does not show the \textit{intrinsic} local TTS, but a sum of variances of several spots on the photocathode and the broadening due to transit time differences. 

The latter effect is also the reason why the TTS measured with a frontal plane wave is significantly larger than the average of the local TTS, as can be seen in figure~\ref{fig:timing:TTS} on the right.


\begin{figure}[tb]
\centering
\includegraphics[scale=0.95]{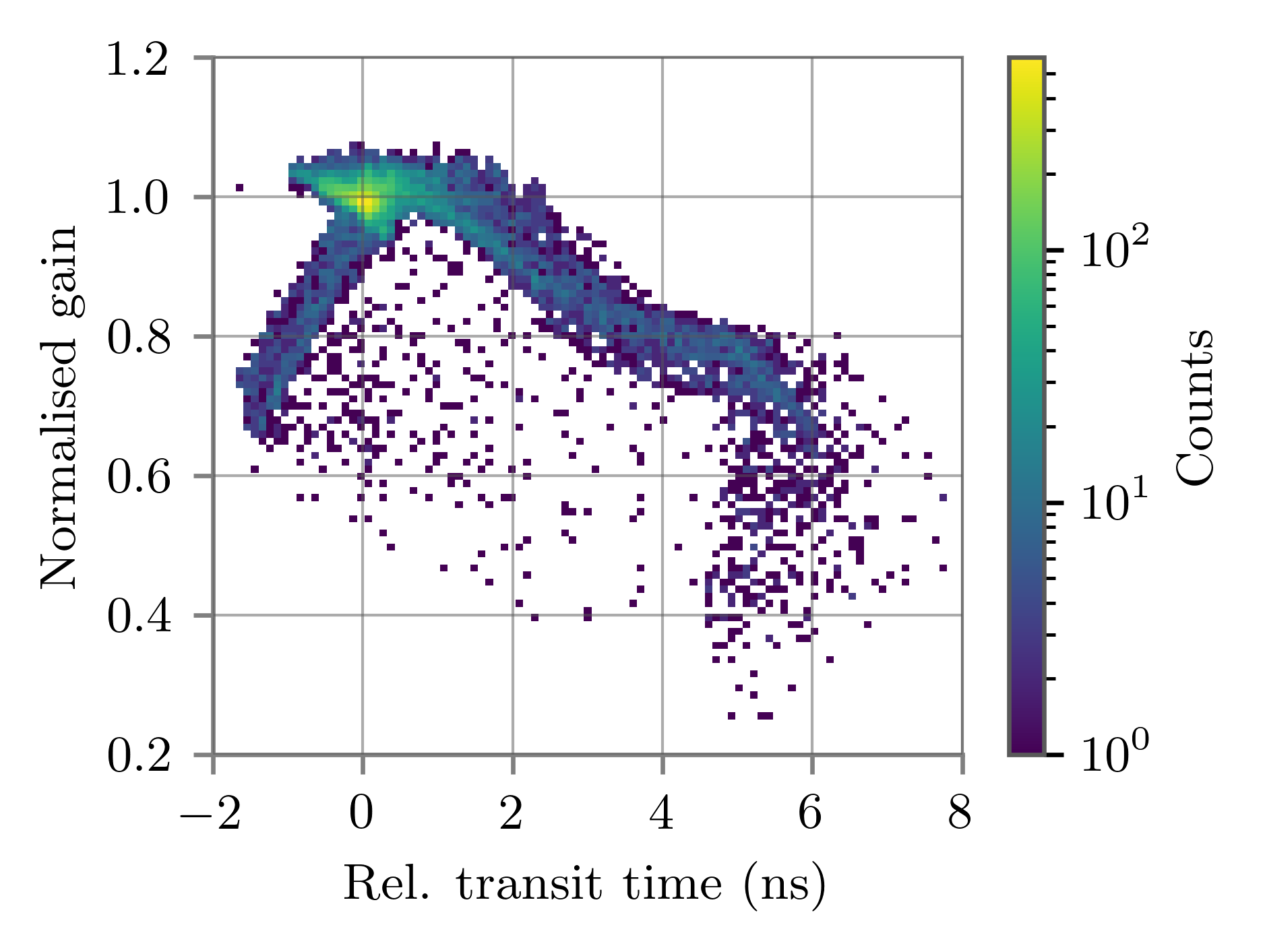}
\hfill
\includegraphics[scale=0.95]{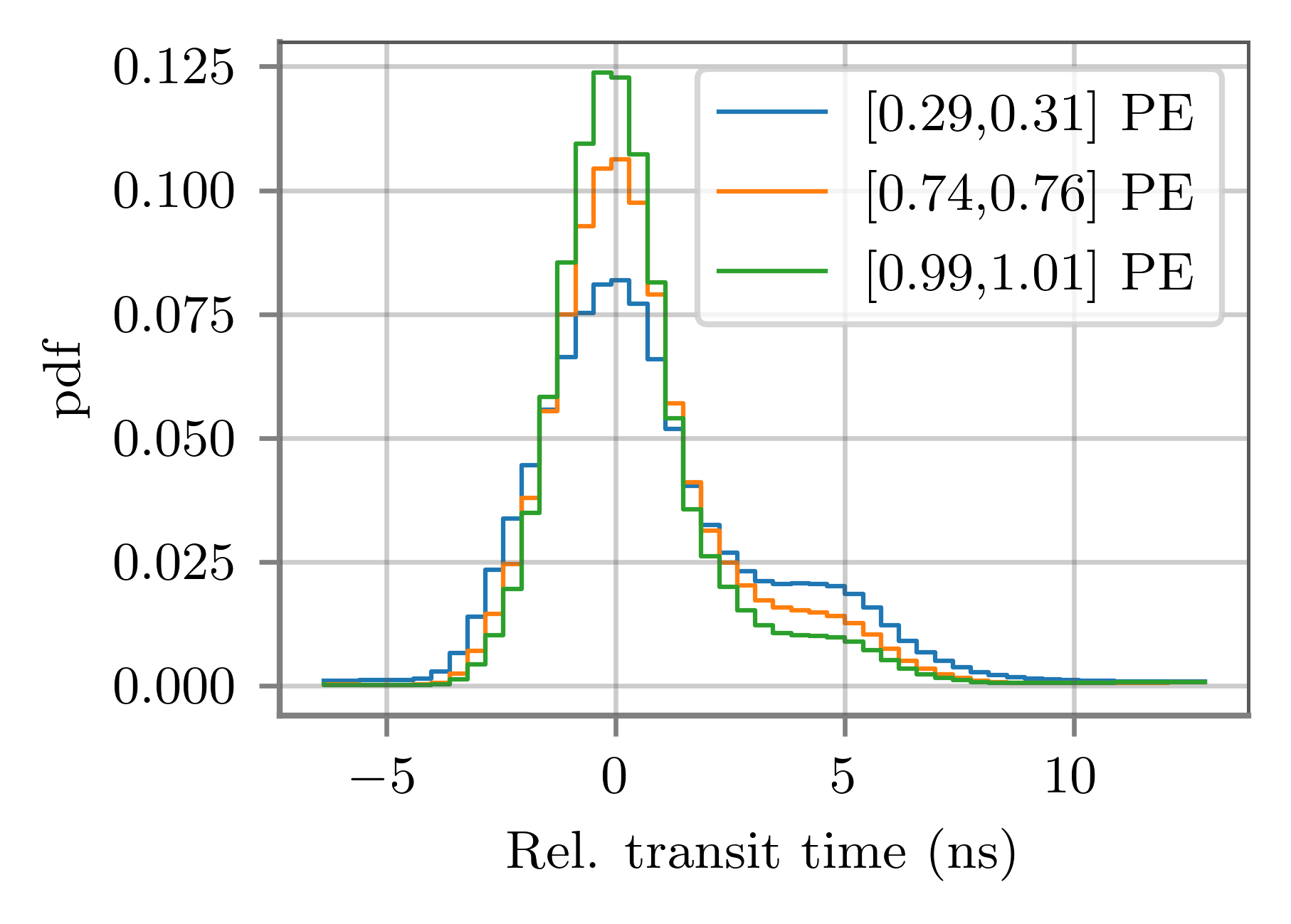}
\caption{\emph{Left}: The relative transit time of all mapped PMTs against the fitted gain on that location. The gain is normalised to the average gain of the central region ($r<\SI{30}{mm}$) of the PMT photocathode. \emph{Right}: Relative transit time distributions for pulses with charge within the [0.29,0.31]$\,$PE, [0.74,0.75]$\,$PE and [0.99,1.01]$\,$PE intervals.
}
\label{fig:corr}
\end{figure}

Comparing the results for transit time and gain, we find that the points on the photocathode with larger transit time deviations also produce pulses with lower charge. This correlation is depicted on the left side of figure~\ref{fig:corr}, where the data of the transit time scan of all PMTs is plotted against the data of the corresponding gain scan normalised to the average gain of the central region of the photocathode ($r<\SI{30}{mm}$). Consequently, the transit time pdf for any given SPE pulse can be constrained with the pulse charge. As an example, the right side of figure~\ref{fig:corr} shows the transit time distributions for pulses within certain charge intervals.

\begin{figure}[tb]
\centering
\includegraphics[scale=0.95]{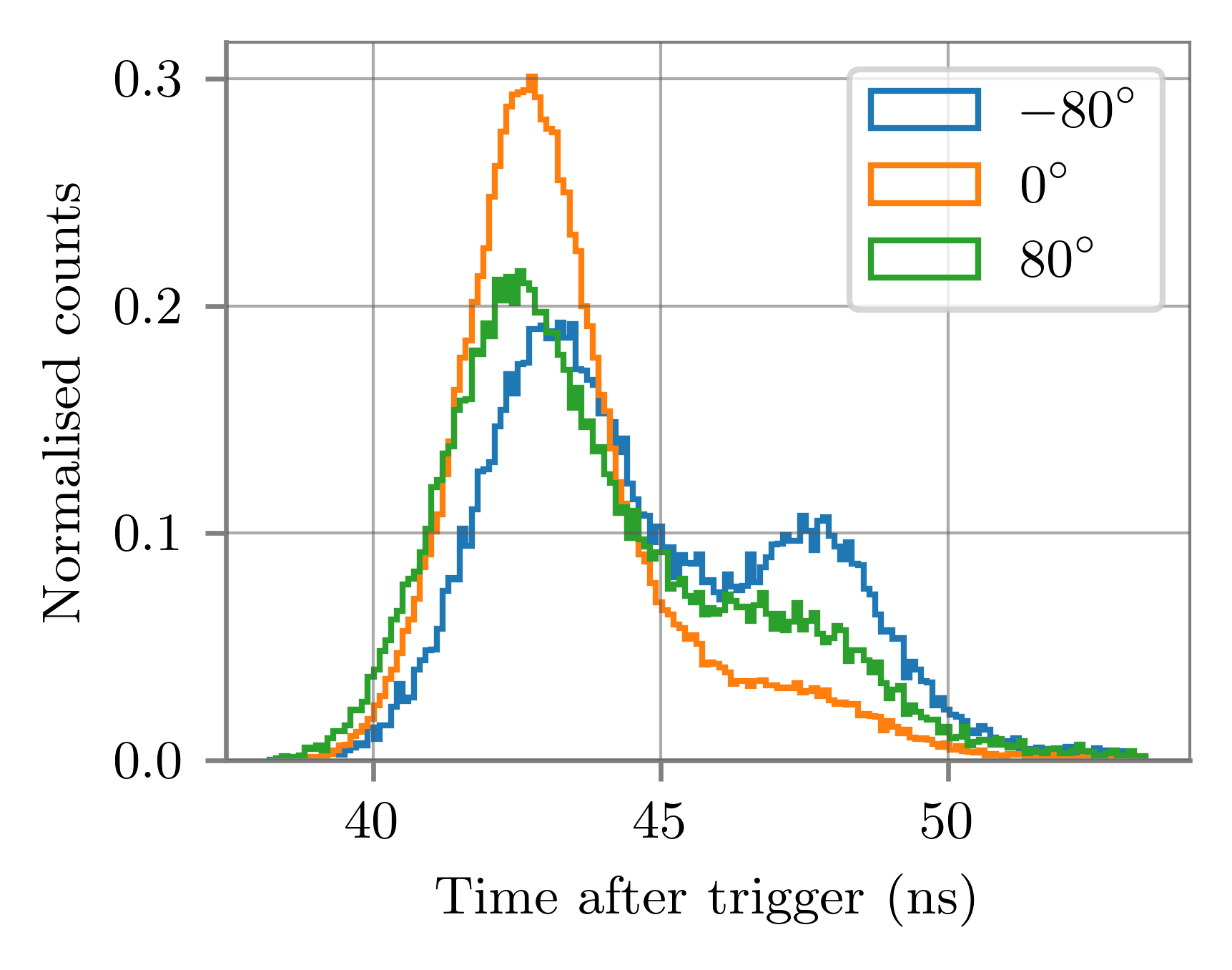}
\caption{Time distribution resulting from a plane wave illuminating the entire photocathode at three different angles to the tube axis. $\SI{0}{\degree}$ corresponds to a plane wave perpendicular to the PMT axis.}
\label{fig:diffangles}
\end{figure}

The transit time deviations for positions on the photocathode with $r>\SI{30}{mm}$ (see figure~\ref{fig:timing:TT} right) are much larger than expected from the width of the distribution (TTS), which means that these systematic shifts will be noticeable in measurements where the whole photocathode is illuminated, giving rise to non-Gaussian distributions. On the left side of figure~\ref{fig:diffangles}, the time distribution for a measurement with frontal illumination of the whole photocathode is presented for three different incidence angles of the plane wave. The variation of the distribution with the angle of incident of the light is due to the changed photon distribution on the photocathode surface.


\subsection{\label{sec:shape}Pulse shape}

\begin{figure}[tb]
\centering
\includegraphics[width=\textwidth]{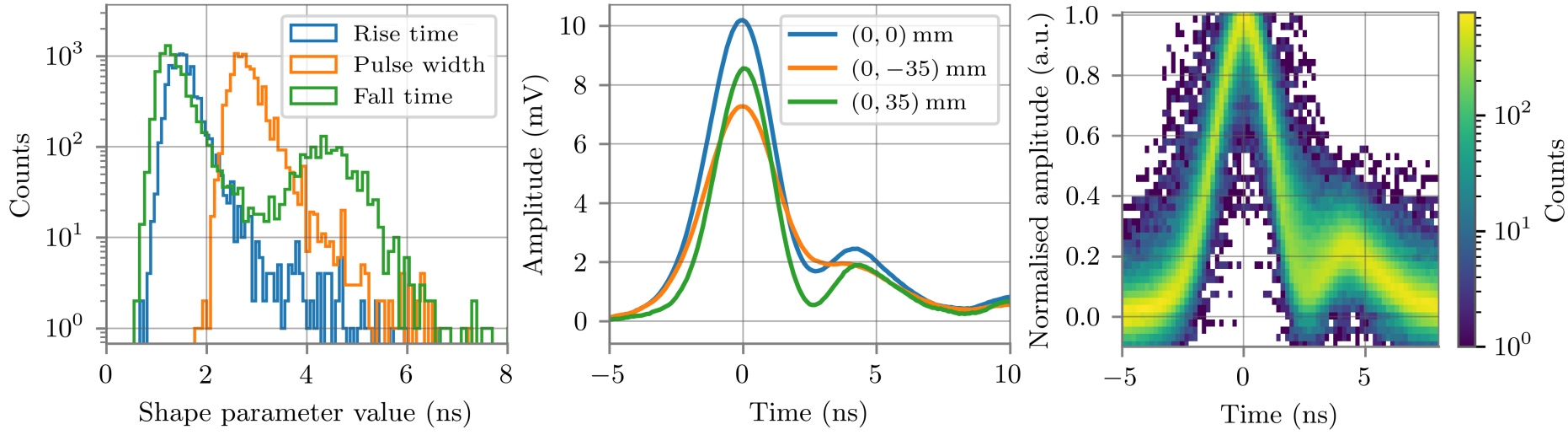}

\caption{\emph{Left}: Distribution of measured shape parameters for a single location on the photocathode. 
\emph{Centre}: Average pulse at the PMT centre, and at two points with different $y$ values representing the extreme cases in the pulse parameter scans. \emph{Right}: Two dimensional histogram of the normalised pulses measured at the PMT centre.}
\label{fig:shape:distaveragePulse}
\end{figure}

\begin{figure}[tb]
\centering
\includegraphics[scale=1]{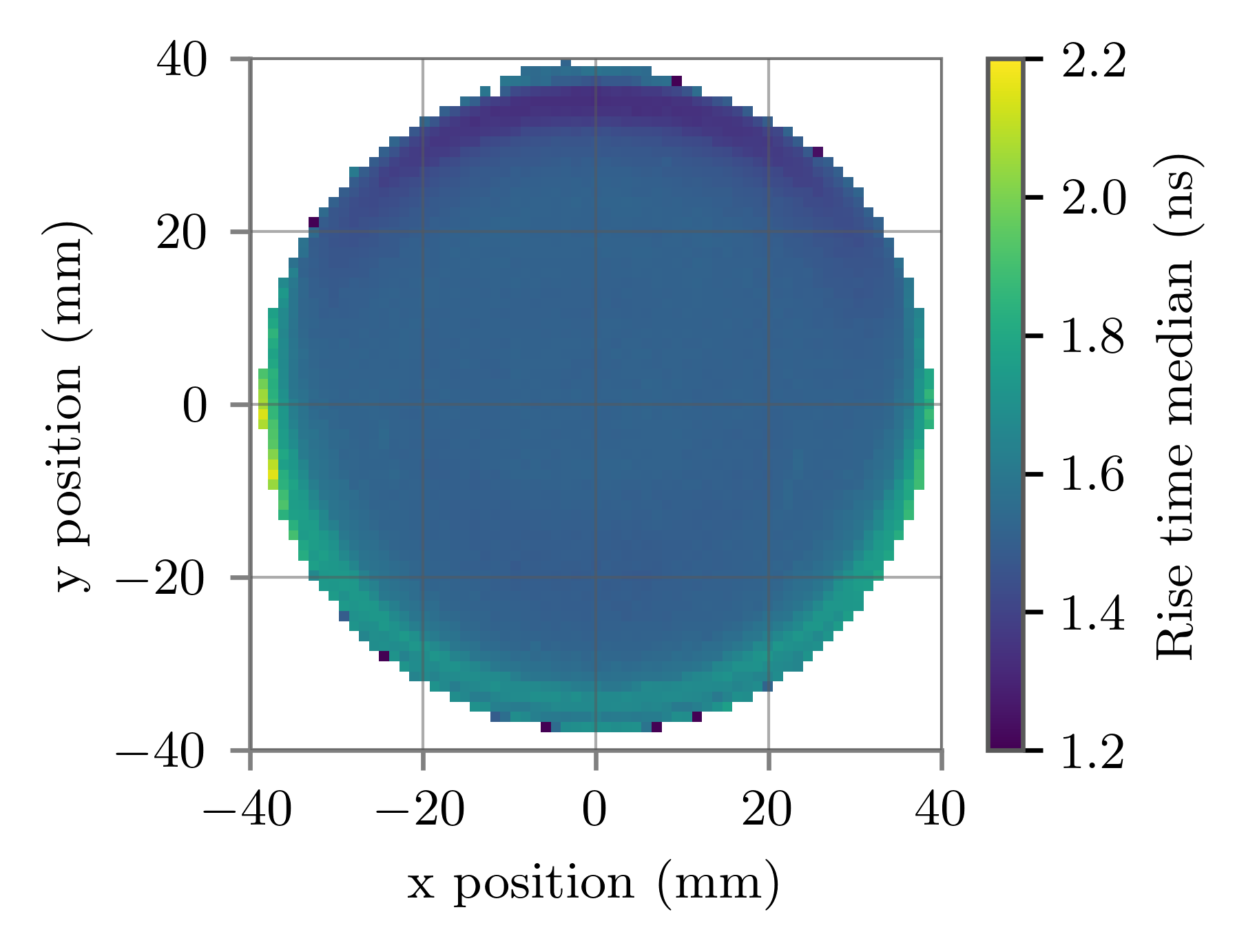}
\hfill
\includegraphics[scale=1]{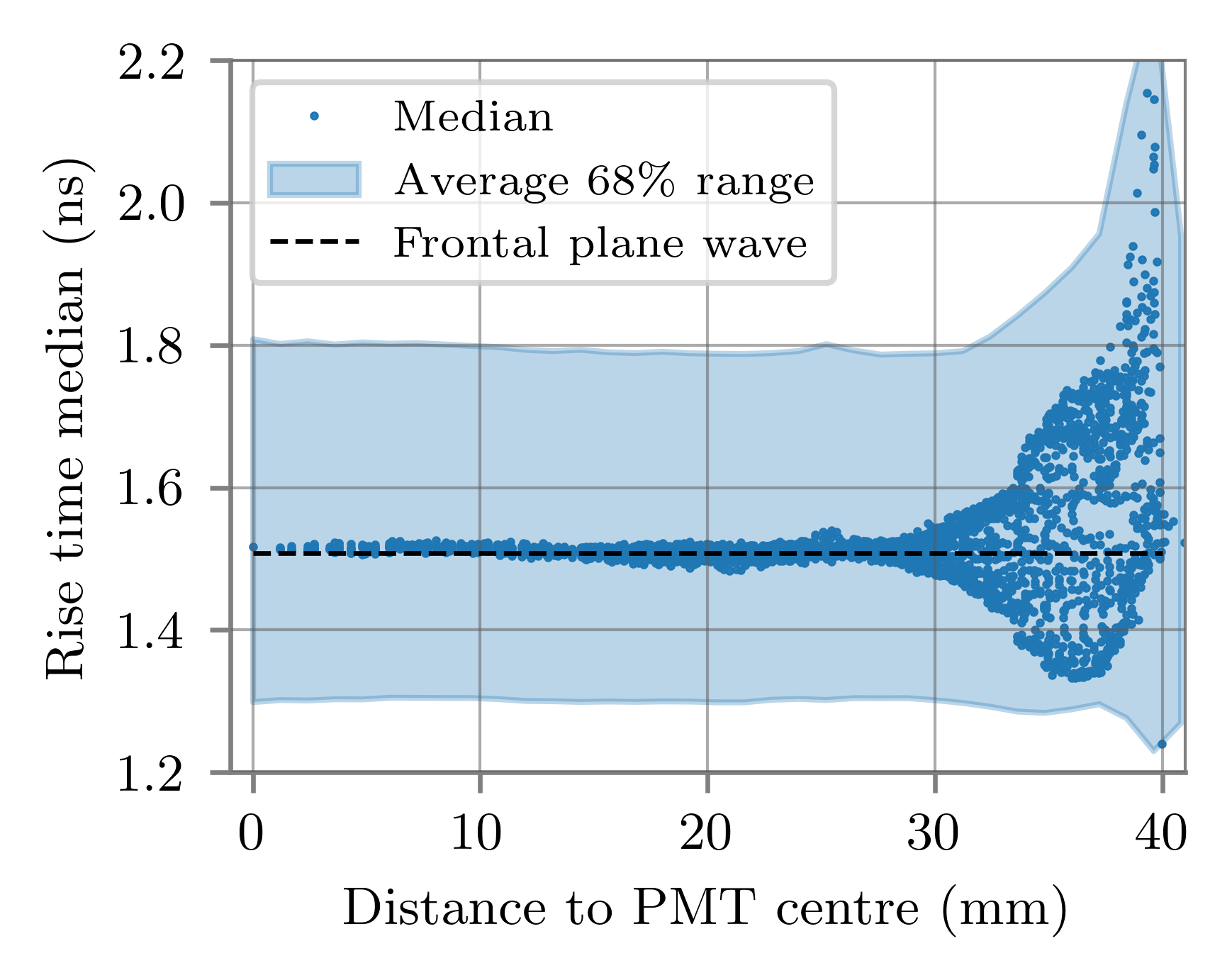}
\caption{\emph{Left}: Scan of the median of the rise-time distribution. \emph{Right}: Median of rise-time against the distance of the measured point to the PMT centre.}
\label{fig:shape:RT}
\end{figure}
\begin{figure}[tb]
\centering
\includegraphics[scale=1]{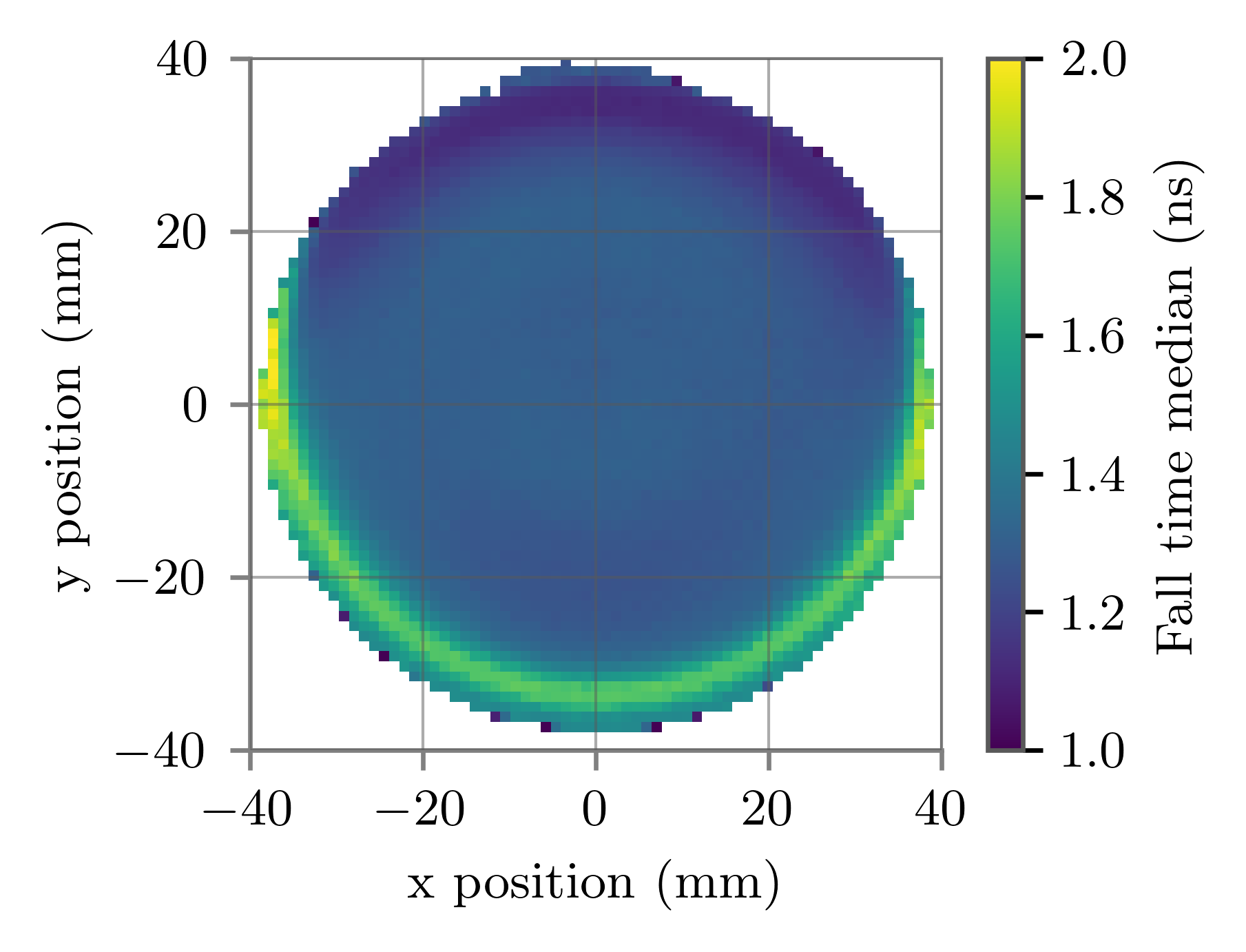}
\hfill
\includegraphics[scale=1]{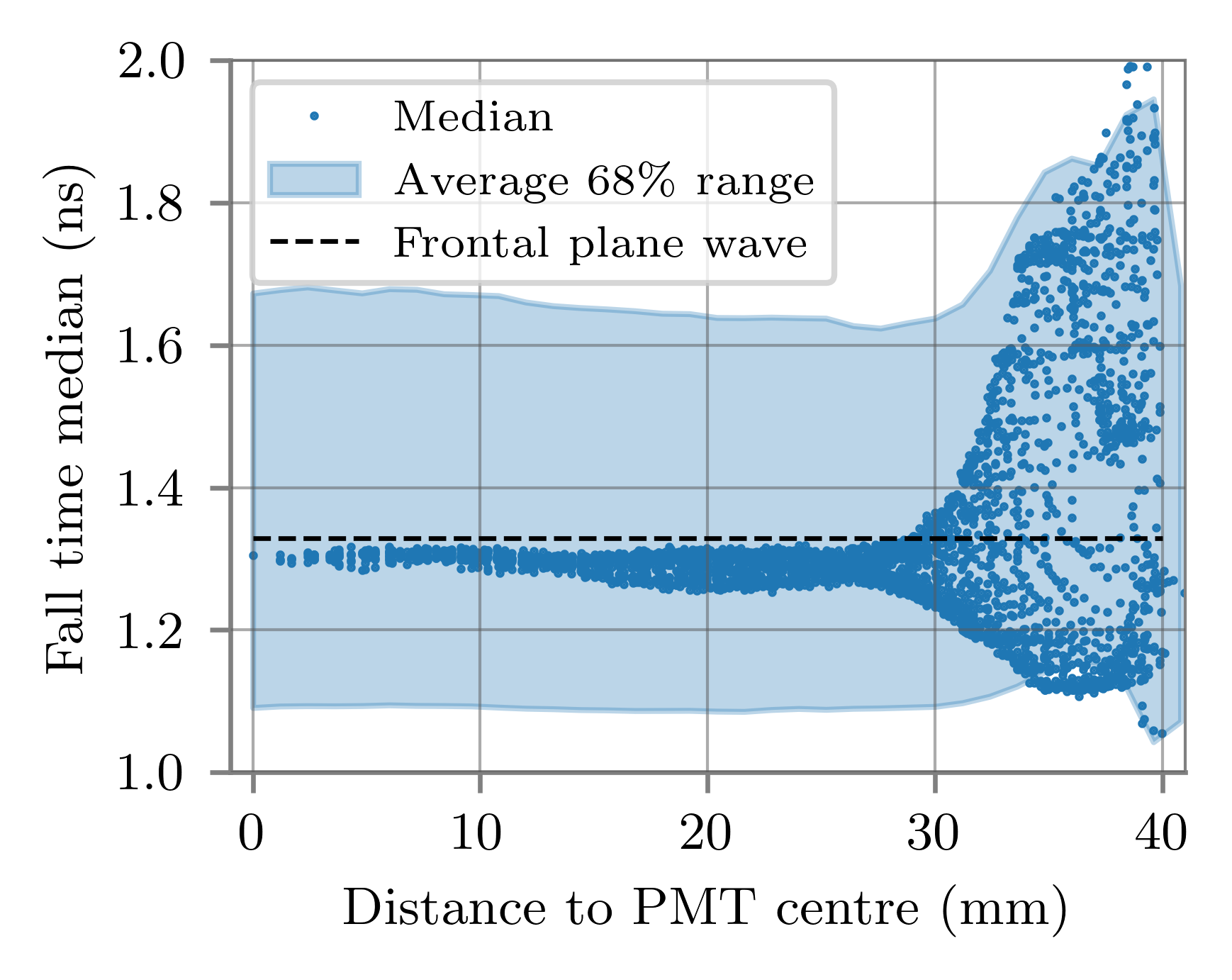}
\caption{\emph{Left}: Scan of the median of the fall-time distribution. \emph{Right}: Median of the fall-time against the distance of the measured point to the PMT centre.}
\label{fig:shape:FT}
\end{figure}
\begin{figure}[tb]
\centering
\includegraphics[scale=1]{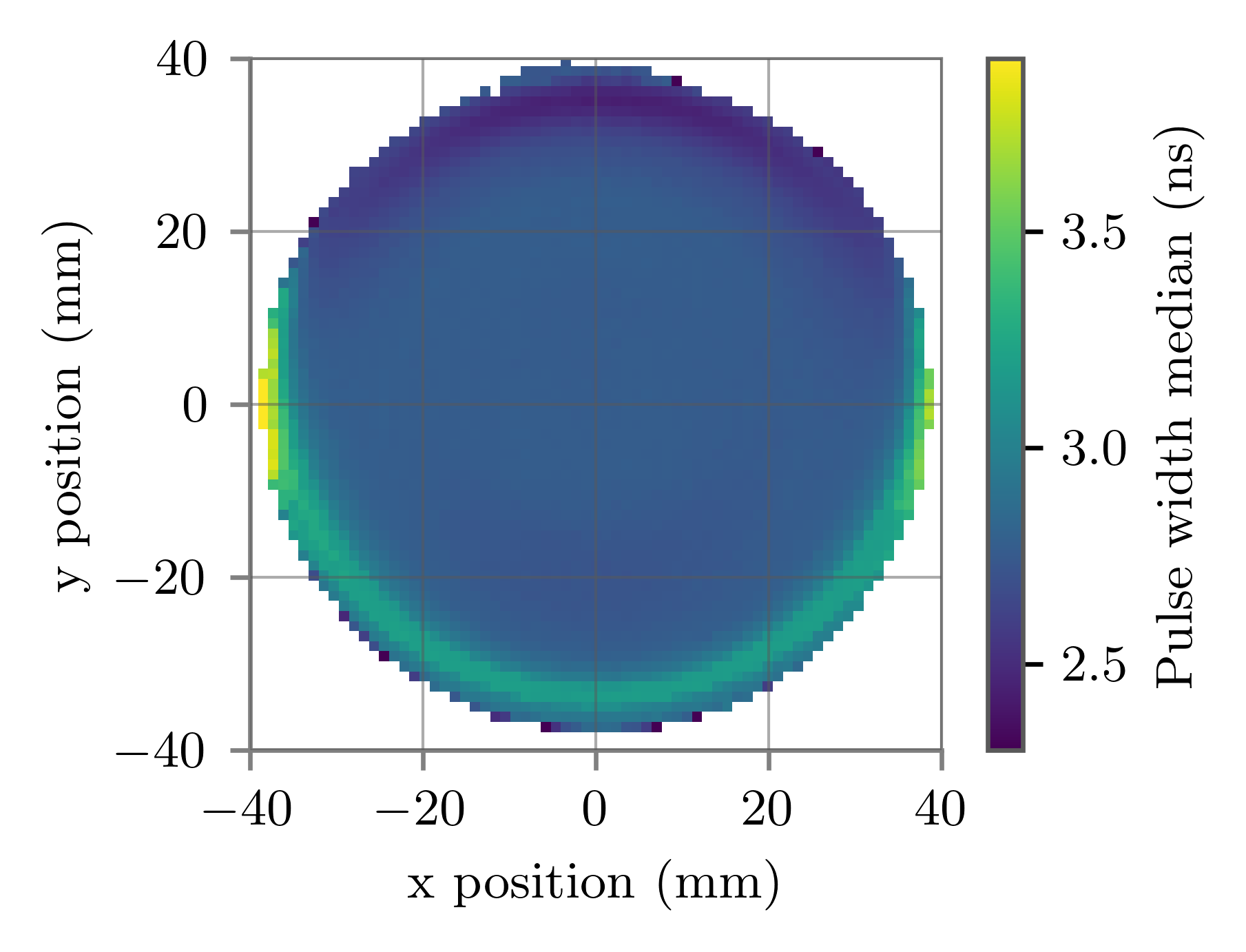}
\hfill
\includegraphics[scale=1]{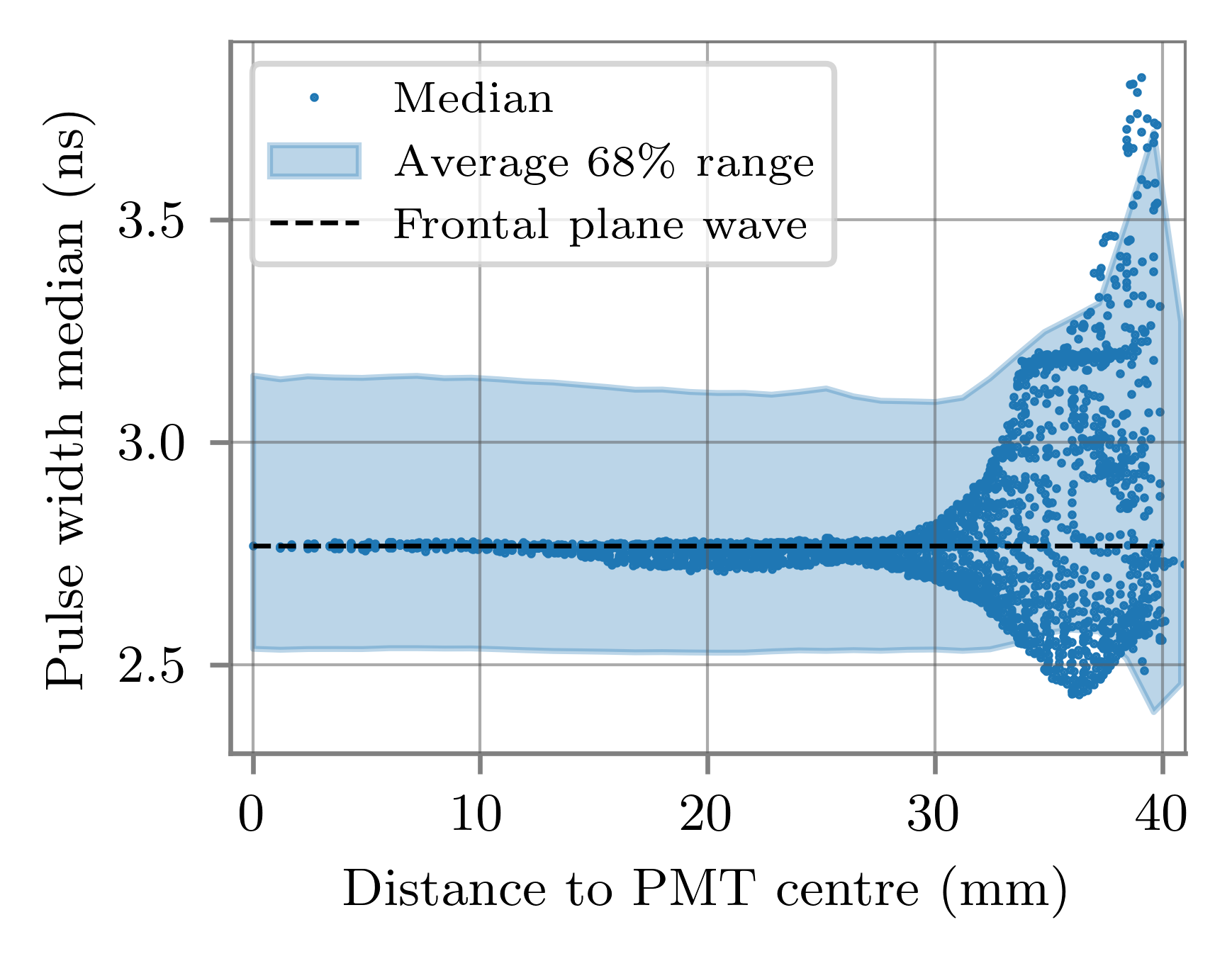}
\caption{\emph{Left}: Scan of the median of the pulse width distribution. \emph{Right}: Median of the pulse width against the distance of the measured point to the PMT centre.}
\label{fig:shape:FW}
\end{figure}

An accurate understanding of the output pulse shape of the PMT is important for the design of the data acquisition electronics of the optical modules and for the waveform reconstruction. As described in section \ref{sec:methods}, the pulse shape parameters of every waveform were saved during the measurement. To limit the analysis to SPE pulses, only waveforms with a measured charge between $Q_1-\sigma_1$ and $Q_1+\sigma_1$ were used. The resulting distributions of pulse width, fall and rise time for the centre position of the PMT \SNPMT\ are depicted in figure~\ref{fig:shape:distaveragePulse}. The pulse width and rise time distributions exhibit a single maximum and are asymmetric, featuring a long tail towards larger values, while the fall time distribution has two local maximums. The maximum at larger values is caused by ringing of the PMT base as discussed in \cite{UnlandElorrieta:2019yhd}. Since this comes from the electronics and not from the PMT itself, values larger than $\SI{3}{ns}$ produced mostly by ringing are excluded in following calculations.

The maps of the median of the shape parameters are depicted on the left side of figure~\ref{fig:shape:RT}, \ref{fig:shape:FT} and \ref{fig:shape:FW} for the rise time, fall time and pulse width, respectively. All three parameters remain relatively constant for the central region of the PMT ($r<\SI{30}{mm}$), while the edges show large deviations. The asymmetry along the $y$-axis is also noticeable for these three parameters, with the largest values at the southern edge of the photocathode ($y\sim-35$\,mm), and shorter pulses at the northern edge ($y\sim35$\,mm).

As in the case of the gain, the distributions of the shape parameters are very wide, as depicted on the right side of figure~\ref{fig:shape:RT}, \ref{fig:shape:FT} and \ref{fig:shape:FW}, where the average $16$th and $84$th percentile, respectively, is plotted against the distance to the PMT centre, together with the median of the distributions. Consequently, photocathode inhomogeneities will not result in pulse changes that are distinguishable from the normal pulse-to-pulse variations. 
The central panel of figure~\ref{fig:shape:distaveragePulse} shows the average pulses from three different positions on the photocathode. Excluding the effect of under-amplification due to a lower local gain, it is noticeable that, although the $(0,-35)\,$mm and $(0,35)\,$mm points show the largest deviation in the pulse parameter scans, the pulses are still quite similar to the one measured at the centre of the PMT, taking into account the intrinsic variance between pulses (right panel in figure~\ref{fig:shape:distaveragePulse}).

Frontal illumination results in values that are very close to the mean value for the central region, since the weightings at the edges largely compensate each other.

\section{Discussion and conclusions}
\label{sec:conclusion}
A homogeneous photocathode response is important for reliable extraction of information from physical events as well as for consistent calibration of PMTs using conventional methods (frontal plane illumination). We find that the central area ($r<\SI{30}{mm}$) of the photocathode of the PMT \pmttype\ is fairly homogeneous for all analysed parameters, but degrades towards the edges. For all investigated parameters an asymmetry along the $y$-axis is observed, which can be attributed to the asymmetry in the PMT geometry introduced by the dynode structure, which is a known effect in studies of spatial uniformity of the output current \cite{Simeonov:99, hamamatsuBook}.

For the gain and pulse shape parameters, the inhomogeneities are small compared to the intrinsic variance of these parameters except for results very close to the edge of the photocathode. The inhomogeneities are small enough that the characterisation of the PMT with frontal plane-wave illumination still provides fairly accurate results. Although the gain in the central region of the PMT is $\sim\SI{6}{\percent}$ larger than that measured with a plane wave, once the entire photocathode is averaged, the measurement with plane waves gives only a $\sim\SI{0.5}{\percent}$ lower value. This is smaller than the systematic bias of the fit model for the charge distribution \cite{osti_1336428}.
 
In contrast, the transit time shows large deviations, with edge-to-edge deviations of several times the TTS. This has three effects: First, if the PMT is characterised with a frontal plane wave, the measured TTS is larger than that at most local positions on the photocathode. Secondly, the time distributions in plane-wave measurements have a non-Gaussian shape. And finally, the non-Gaussian shape of the distribution depends on the angle of incidence of the plane wave.
 
The large transit time deviations may affect the reconstruction algorithms used for neutrino events, especially in cases of low-light, since the time distribution is usually modelled as a single Gaussian with a TTS measured with frontal plane waves. This should be studied further, especially for reconstructions in detectors deployed in water, since the prompt light distributions are very narrow due to the large scattering length in this medium \cite{Adri_n_Mart_nez_2016}. Ice, on the other hand, has a shorter scattering length, and the direction of the detected photons tends to be more isotropic, widening the time distribution of the detected photons. Furthermore, in events with large amounts of detected photons, the use of multi-PMT optical modules substantially counteracts the effects of transit time inhomogeneities, as these are detected in several PMTs pointing at different directions, reducing a possible bias due to the position of the light source with respect to the modules.

Positions on the photocathode that result in large transit-time deviations also lead to under-amplified pulses on average, as seen in figure~\ref{fig:corr}. It should be studied whether this correlation can be exploited in pulse reconstruction algorithms, since the PDF of the transit time of any given pulse can be constrained by its charge (see right side of figure~\ref{fig:corr}). However, since the PDFs for different charges are relatively similar, the improvement is likely to be small.

\acknowledgments

This work was supported by the German Bundesministerium f\"ur Bildung und Forschung (BMBF) Verbundforschung grants 05A20PM2.

\appendix
\section{Transit time spread correction for MPE Poissonian light distributions}
\label{appendix:TTscorrection}

\begin{figure}[htbp]
\centering
\includegraphics[scale=1]{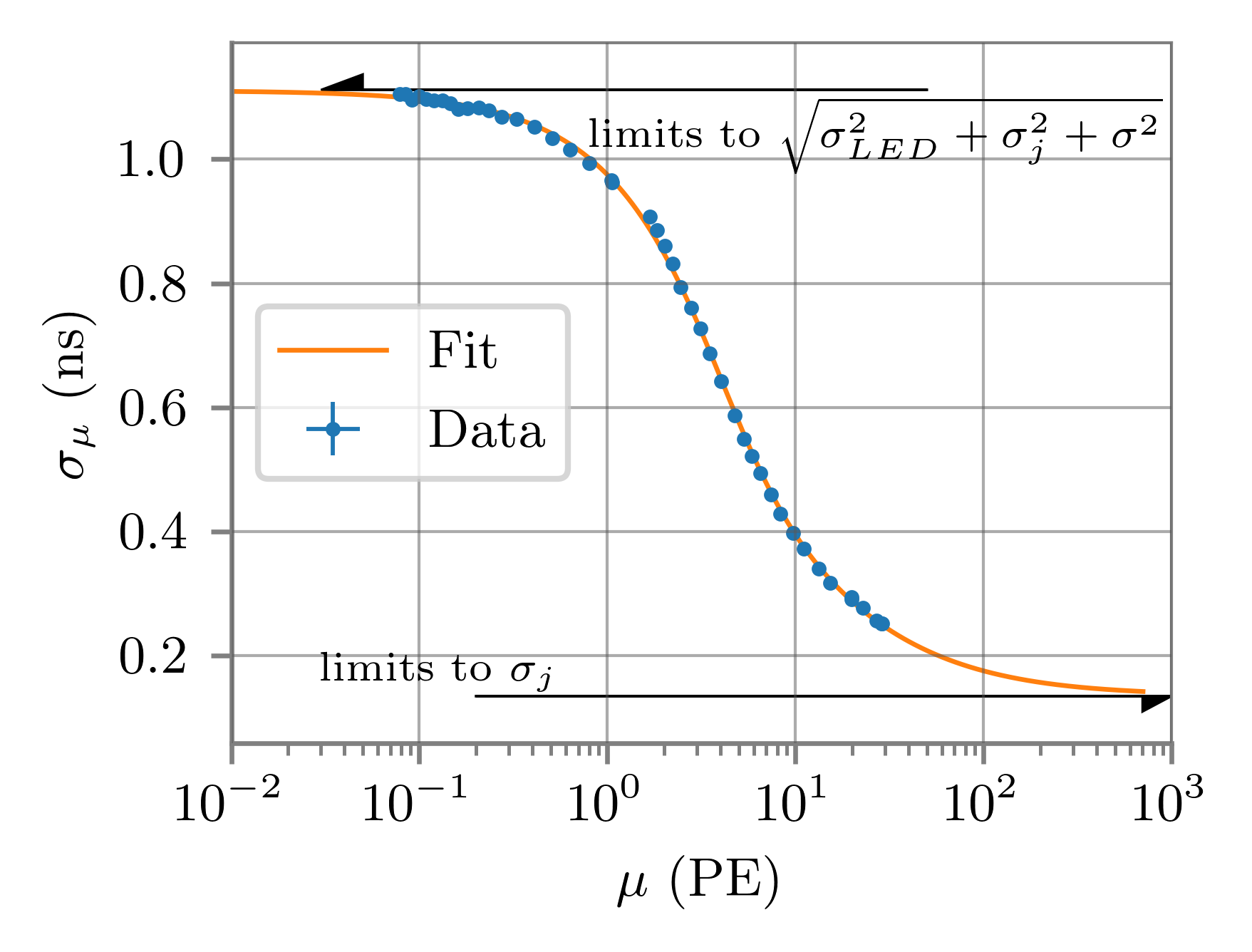}
\hfill
\includegraphics[scale=1]{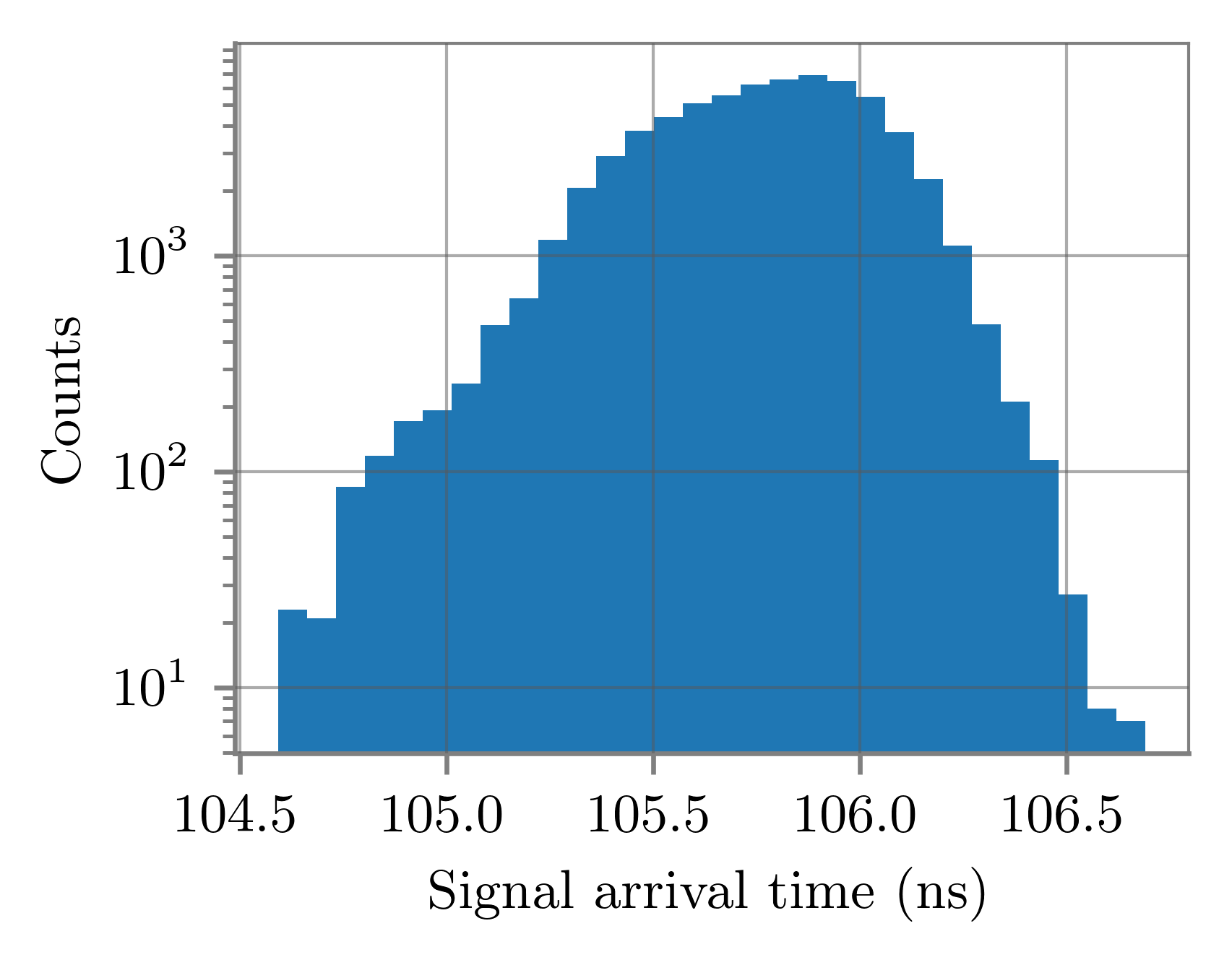}
\caption{\emph{Left}: Width of the transit time distribution against the mean number of detected photons. The data was fitted with equation~\ref{app:ttseq}. \emph{Right}: Distribution of the relative LED emission time measured with a fast SPAD.}
\label{fig:appendix:mucorrection}
\end{figure}

Typically, the TTS of a PMT is measured at low light intensities, which ensures that the PMT detects almost only SPEs. The time distributions of MPE pulses are narrower due to statistical effects:
assuming a variance of TTs of different SPEs $\rm{Var}(\rm{TT})= \sigma^{2}$, the variance of the mean transit time $\overline{\textrm{TT}}$ of $n$ photoelectrons is as follows
\begin{equation}
\textrm{Var}\big[\overline{\textrm{TT}}\big]=\textrm{Var}\Big[\frac{1}{n}\sum^{n}_{i=1}\textrm{TT}_i\Big]=\frac{1}{n^2}\sum^{n}_{i=1}\textrm{Var}[\textrm{TT}_i]= \frac{\sigma^2}{n}.  
\end{equation}
Therefore, the measured TTS of such a distribution would be a factor of $\frac{1}{\sqrt{n}}$ smaller than the SPE-TTS. 
In real measurements, the time distribution is a convolution between the time distribution of an SPE and the distribution of the number of detected photons. The former can be assumed as a Gaussian $G(t, TT,\hat{\sigma})$, where $\hat{\sigma} = \sqrt{\sigma^2+\sigma^2_{\textrm{LED}}}$ is the sum of variances of the transit time distribution and the LED light time profile. In the case of an LED, the number of detected photons $n$ follows Poisson statistics $P(n,\mu)$, with the mean $\mu$. Thus, the pdf of the transit time $f(t)$ is
\begin{equation}
    f(t) = \frac{1}{N} \cdot \sum_{n = 1}^{\infty}P(n,\mu)\cdot G(t, \textrm{TT}, \hat{\sigma}/\sqrt{n}),
\end{equation}
with $N=(1-e^{-\mu})^{-1}$ a normalisation factor, since the case of zero photons detected does not contribute to the distribution. The variance of this pdf is calculated with:
\begin{equation} \label{eq1}
\begin{split}
\sigma^2_\mu=\textrm{Var}[f(t)] & = -\textrm{E}[t]^2 + \int t^2 \cdot f(t) dt\\
 & =  \frac{\hat{\sigma}^2 }{e^{\mu}-1} [\rm{Ei}(\mu)-\gamma - \log (\mu)],
\end{split}
\end{equation}
where $\textrm{E}[t] = \int t \cdot f(t) dt = \textrm{TT} $ is the expected value, $\rm{Ei}(\mu)=\int_{-\infty}^{\mu} \textrm{e}^{x}\cdot x^{-1} dx$ the exponential integral and $\gamma$ the Euler-Mascheroni constant.

Finally, there is the constant contribution of the electronic jitter $\sigma_{\textrm{j}}$ from the LED driver and the trigger, which is independent of the number of measured photons. Therefore, the measured standard deviation $\sigma_{\mu}$ for Poissonian light distributions with a mean number of detected photons $\mu$ is described by
\begin{equation}
\label{app:ttseq}
    \sigma_{\mu}(\mu) = \sqrt{\frac{(\sigma^2+\sigma^2_{\textrm{LED}})}{e^{\mu}-1} [\rm{Ei}(\mu)-\gamma - \log (\mu)] + \sigma^2_{j}}.
\end{equation}

Solving for $\sigma$, it is possible to calculate the ``real'' TTS for SPEs.

One of the PMTs was measured with light intensities in a range of three orders of magnitude and the measured standard deviation was fitted with the equation~\ref{app:ttseq}. Since $\sigma_{\mu}$ limits to $\sigma_{\textrm{j}}$ as $\mu\rightarrow\infty$, it is possible to extract the electronic jitter from the fit, which is determined to be $\SI{136\pm6}{ps}$. Since there is a degeneracy between the SPE TTS $\sigma$ and the LED contribution $\sigma_{\textrm{LED}}$, it is necessary to determine the latter with an independent measurement.

The LED time variance was determined measuring the light output with a fast avalanche photodiode\footnote{IDQ ID100}, which has a time resolution (FWHM) of $\SI{<60}{ps}$. The arrival time distribution is depicted on the right side of figure~\ref{fig:appendix:mucorrection}. The distribution is almost Gaussian with a tail towards the shorter values and has a standard deviation $\sigma_{\textrm{LED,j}}=\SI{271\pm1}{ps}$. The width of the time profile of the LED is therefore $\sigma_{\textrm{LED}}=\sqrt{\sigma_{\textrm{LED,j}}^2-\sigma_{\textrm{j}}^2}=\SI{234\pm4}{ps}$.

\bibliographystyle{JHEP}
\bibliography{bib.bib}

\providecommand{\href}[2]{#2}\begingroup\raggedright\begin{thebibliography}{10}

\bibitem{Adri_n_Mart_nez_2016}
S.~Adri{\'{a}}n-Mart{\'{\i}}nez, M.~Ageron, F.~Aharonian, S.~Aiello, A.~Albert,
  F.~Ameli et~al., \emph{Letter of intent for {KM3NeT} 2.0},
  \href{https://doi.org/10.1088/0954-3899/43/8/084001}{\emph{Journal of Physics
  G: Nuclear and Particle Physics} {\bfseries 43} (2016) 084001}.

\bibitem{ishihara2019icecube}
A.~Ishihara, \emph{The icecube upgrade -- design and science goals},
  \href{https://arxiv.org/abs/1908.09441}{{\ttfamily 1908.09441}}.

\bibitem{Abe:2018uyc}
{\scshape Hyper-Kamiokande} collaboration, \emph{{Hyper-Kamiokande Design
  Report}},  \href{https://arxiv.org/abs/1805.04163}{{\ttfamily 1805.04163}}.

\bibitem{nim:a718:513}
{\scshape KM3NeT} collaboration, \emph{{The multi-PMT optical module for
  KM3NeT}}, \href{https://doi.org/10.1016/j.nima.2012.11.049}{\emph{Nucl.\
  Inst.\ Meth.} {\bfseries A718} (2013) 513}.

\bibitem{Classen:2019/w}
L.~Classen, C.~Dorn, A.~Kappes, T.~Karg, M.~Kossatz, A.~Kretzschmann et~al.,
  \emph{{A multi-PMT Optical Module for the IceCube Upgrade}},  in
  \emph{Proceedings of 36th International Cosmic Ray Conference {\textemdash}
  PoS(ICRC2019)}, vol.~358, p.~855, 2019,
  \href{https://doi.org/10.22323/1.358.0855}{DOI}.

\bibitem{DeRosa:2020wbm}
{\scshape Hyper-Kamiokande} collaboration, \emph{{A multi-PMT photodetector
  system for the Hyper-Kamiokande experiment}},
  \href{https://doi.org/10.22323/1.390.0831}{\emph{PoS} {\bfseries ICHEP2020}
  (2021) 831}.

\bibitem{UnlandElorrieta:2019yhd}
M.A.~Unland~Elorrieta, L.~Classen, J.~Reubelt, S.~Schmiemann, J.~Schneider and
  A.~Kappes, \emph{{Characterisation of the Hamamatsu R12199-01 HA MOD
  photomultiplier tube for low temperature applications}},
  \href{https://doi.org/10.1088/1748-0221/14/03/P03015}{\emph{JINST} {\bfseries
  14} (2019) P03015} [\href{https://arxiv.org/abs/1902.01714}{{\ttfamily
  1902.01714}}].

\bibitem{martin}
{M. A. Unland Elorrieta}, \emph{Studies on dark rates induced by radioactive
  decays of the multi-PMT digital optical module for future IceCube
  extensions}, master thesis, Institut f\"ur Kernphysik, Westf\"alische
  Wilhelms-Universit\"at M\"unster, 2017.

\bibitem{Bellamy1994}
E.~Bellamy, G.~Bellettini, J.~Budagov, F.~Cervelli, I.~Chirikov-Zorin,
  M.~Incagli et~al., \emph{Absolute calibration and monitoring of a
  spectrometric channel using a photomultiplier},
  \href{https://doi.org/10.1016/0168-9002(94)90183-x}{\emph{Nucl.\ Inst.\
  Meth.} {\bfseries A339} (1994) 468}.

\bibitem{Wright2017}
A.G.~Wright, \emph{The Photomultiplier Handbook}, Oxford University Press
  (2017),
  \href{https://doi.org/10.1093/oso/9780199565092.001.0001}{10.1093/oso/9780199565092.001.0001}.

\bibitem{Simeonov:99}
V.~Simeonov, G.~Larcheveque, P.~Quaglia, H.~van~den Bergh and B.~Calpini,
  \emph{Influence of the photomultiplier tube spatial uniformity on lidar
  signals}, \href{https://doi.org/10.1364/AO.38.005186}{\emph{Appl. Opt.}
  {\bfseries 38} (1999) 5186}.

\bibitem{hamamatsuBook}
{HAMAMATSU PHOTONICS K. K.}, \emph{Photomultiplier tubes - Basics and
  Applications}, Hamamatsu Photonics K.K., 4~ed. (2017).

\bibitem{osti_1336428}
R.~Saldanha, L.~Grandi, Y.~Guardincerri and T.~Wester, \emph{Model independent
  approach to the single photoelectron calibration of photomultiplier tubes},
  \href{https://doi.org/10.1016/j.nima.2017.02.086}{\emph{Nuclear Instruments
  and Methods in Physics Research. Section A, Accelerators, Spectrometers,
  Detectors and Associated Equipment} {\bfseries 863} (2017) }.

\end{thebibliography}\endgroup

\end{document}